\documentclass[aps,pra,twocolumn,showpacs,10pt,amsmath,footinbib]{revtex4-1}
\usepackage{graphicx}
\usepackage[usenames,dvipsnames]{xcolor}
\usepackage{braket}
\usepackage{array}
\usepackage{amsmath}
\DeclareMathOperator*{\argmax}{arg\,max}

\usepackage{tikz}
\usepackage{pgfplots}

\newcommand{\pa}{\partial}

\def\gappeq{\mathrel{ \rlap{\raise.5ex\hbox{$>$}}
                      {\lower.5ex\hbox{$\sim$}} } }
\def\lappeq{\mathrel{ \rlap{\raise.5ex\hbox{$<$}}
                      {\lower.5ex\hbox{$\sim$}} } }
                      
                      \newcommand{\del}[1]{\textcolor{red}{}}
                      
\begin{document}

\title{Kelvin-Helmholtz instability in an atomic superfluid}

\author{A. W. Baggaley}\email{andrew.baggaley@ncl.ac.uk}
\affiliation{Joint Quantum Centre (JQC) Durham--Newcastle, School of Mathematics and Statistics, Newcastle University, Newcastle upon Tyne, NE1 7RU, United Kingdom}
\author{N. G. Parker}
\affiliation{Joint Quantum Centre (JQC) Durham--Newcastle, School of Mathematics and Statistics, Newcastle University, Newcastle upon Tyne, NE1 7RU, United Kingdom}

\date{\today}

\begin{abstract}
We demonstrate an experimentally feasible method for generating the classical Kelvin-Helmholtz instability in a single component atomic Bose-Einstein condensate.
By progressively reducing a potential barrier between two counter-flowing channels we seed a line of quantised vortices, which precede to form progressively larger clusters, mimicing the classical roll-up behaviour of the Kelvin-Helmholtz instability. 
This cluster formation leads to an effective superfluid shear layer, formed through the collective motion of many quantised vortices. 
From this we demonstrate a straightforward method to measure the effective viscosity of a turbulent quantum fluid in a system with a moderate number of vortices, within the range of current experimental capabilities.
\end{abstract}

\maketitle

\section*{Introduction}

The hydrodynamic instabilities of ordinary viscous fluids are a cornerstone of fluid mechanics, governing the breakdown of laminar flow and the transition to turbulence \cite{DrazinReid,Charru}, and of great importance across fluid motion in engineering, meteorology, oceanography, astrophysics and geophysics.   The modern era has seen the advent of superfluids, realised in the laboratory in the form of superfluid Helium \cite{Leggett1999}, ultracold atomic gases (Bose-Einstein condensates (BECs) \cite{Dalfovo1999}  and degenerate Fermi gases \cite{Giorgini2008}) and quantum fluids of light \cite{Carusotto2013}.  The macroscopic quantum behaviour leads to several key distinctions from ordinary fluids \cite{Tsubota2013b}.  Firstly, viscosity is absent in the quantum fluid.  Secondly, when the fluid velocity exceeds a critical magnitude the flow is dissipated through elementary excitations.  Thirdly, vorticity is constrained to exist only a discrete filaments with quantised vorticity.  Given these deep apparent differences, an ongoing direction of research is to establish whether the paradigm instabilities of ordinary fluids have analogs in superfluids, how they manifest and in what ways they are similar \cite{Tsubota2013}.  Considerable attention has been given to the instability of laminar superfluid flow past obstacles and surfaces, revealing quantum analogs of the classical wakes including the von K\'arm\'an vortex street \cite{Sasaki2010,Stagg2014,Shin2016} and the boundary layer \cite{Stagg2017}.   Systems of two immiscible BECs are predicted to exhibit the Rayleigh-Taylor instability of the interface between them \cite{Sasaki2009, Gautum2010,Bezett2010,Kobyakov2011,Jia2012,Kadokura2012, Kobyakov2014}.    Meanwhile, the presence of magnetic dipolar atomic interactions leads to instabilities analogous to those found in ferrofluids, including the Rosenweig instability \cite{Saito2009,Kadau2016} and fingering instability \cite{Xi}.

The Kelvin-Helmholtz (KH) instability is one of the most elementary hydrodynamic instabilities, first formulated by Helmholtz \cite{Helmholtz} and Kelvin \cite{Kelvin} in the nineteenth century, and describes the instability of the interface between two parallel fluid streams with different velocities.  Under suitable conditions, the interface undergoes a dynamical instability characterised by exponential growth of perturbations.  The interface tends to roll up, destroying the steady laminar flow, and often initiating a transition to turbulence.  The simplest flow which supports the KH instability is for two streams within a single inviscid incompressible fluid, for which the instability occurs for all values of the relative speed.   The KH instability also arises for two streams of different fluids with different densities, in which case the KH instability can become superposed by the buoyancy-driven Rayleigh-Taylor instability \cite{Charru}.  

To date, superfluid analogs of the KH instability have been considered at the interface of two distinct superfluids.  The KH instability between the A and B phase of superfluid $^3$He has been detected experimentally under rotation \cite{Blaauwgeers} and analysed theoretically \cite{Volovik1,Volovik2,Finne}.   The KH instability between nuclear superfluids in a neutron star has been proposed as the trigger for pulsar glitches \cite{Mastrano}.  It has also been discussed at the interfaces between the normal fluid and superfluid \cite{Henn,Korshunov}, between $^3$He and $^4$He \cite{Burmistrov}, and the interface between two components of an immiscible binary BEC \cite{Takeuchi,Suzuki,Kobyakov2014}.  In these cases, the presence of two distinct fluids complicates the behaviour, including buoyancy effects \cite{Kobyakov2014} and a crossover to a counterflow instability if there is significant overlap of the fluids at the interface \cite{Suzuki}.  

The goal of this paper is to demonstrate that the KH instability can be realized within a single component superfluid and that this  prototypical incarnation of the KH instability is achieveable with current experimental technologies.  We will also see that the KH instability leads to the formation of vortex clusters which, when coarse-grained, mimic a viscous shear layer.  This facilitates a measurement of the effective viscosity in a system with a moderate number of vortices, well within the limits of current experimental systems.

\section*{Model}

We model a weakly-interacting atomic superfluid BEC in two-dimensions through a macroscopic wavefunction $\Psi(x,y,t)$ which evolves according to the Gross-Pitaevskii equation (GPE) \cite{Pethick,Pitaevskii,Barenghi},
\begin{equation}
i  \hbar \frac{\pa\Psi }{\pa t}=\left[
-\frac{\hbar^2}{2m}\left(\frac{\partial^2}{\partial x^2} + \frac{\partial^2}{\partial y^2} \right) 
+V(x,y,t)+g\left|\Psi \right|^{2}\right]\Psi.
\label{eq:GPE}
\end{equation}
Here $m$ is the atomic mass and $g$ is a nonlinear coefficient arising from the contact-like atomic interactions.  As is typical of most BEC experiments (to guarantee the stability of the BEC against collapse), we consider repulsive atomic interactions, $g>0$. The 2D atomic density follows from the wavefunction as $n(x,y,t)=|\Psi(x,y,t)|^2$ and the fluid velocity as ${\bf v}(x,y,t)=(\hbar/m)\nabla \theta(x,y,t)$, where $\theta(x,y,t)$ is the phase distribution of $\Psi$.  Time-independent solutions of the GPE satsify $i \hbar {\Psi}_t= \mu \Psi$, where $\mu$ is the chemical potential of the condensate.  Advanced techniques using optical and magnetic fields now allow for almost arbitrary spatial and temporal control over the external potential $V$ experienced by the atoms \cite{Henderson}.  

We non-dimensionalise the GPE based on ``natural units" \cite{Barenghi} in which the unit of length is the healing length $\xi=\hbar/\sqrt{m \mu}$, the unit of time is $\hbar/\mu$, the unit of energy is $\mu$, and the unit of density is $n_0=\mu/g$.  The corresponding unit of speed is the speed of sound $c=\sqrt{\mu/m}$.  
We proceed by performing numerical simulations of the GPE (non-dimensionalised using the above units). We  consider
the domain $-D_x \leq x \leq  D_x$, $-D_y \leq y \leq  D_y$,
with $D_x=128$ and $D_y=64 $. Space is discretized
 onto a $N_x \times N_y=512 \times 256$ uniform cartesian mesh, spatial derivatives are approximated by a $6^{\rm th}$--order finite difference scheme and a $3^{\rm rd}$--order Runge-Kutta scheme is used for time evolution, with time-step $\delta t = 5 \times 10^{-3}$. Periodic boundaries are taken in the $x$ (streamwise) direction and zero boundaries in the $y$ (transverse) direction, although our choice of potential means our system is effectively independent of the choice of boundary conditions in the transverse direction.
 
\begin{figure}[t]
\centering%
\includegraphics[width=0.8\linewidth]{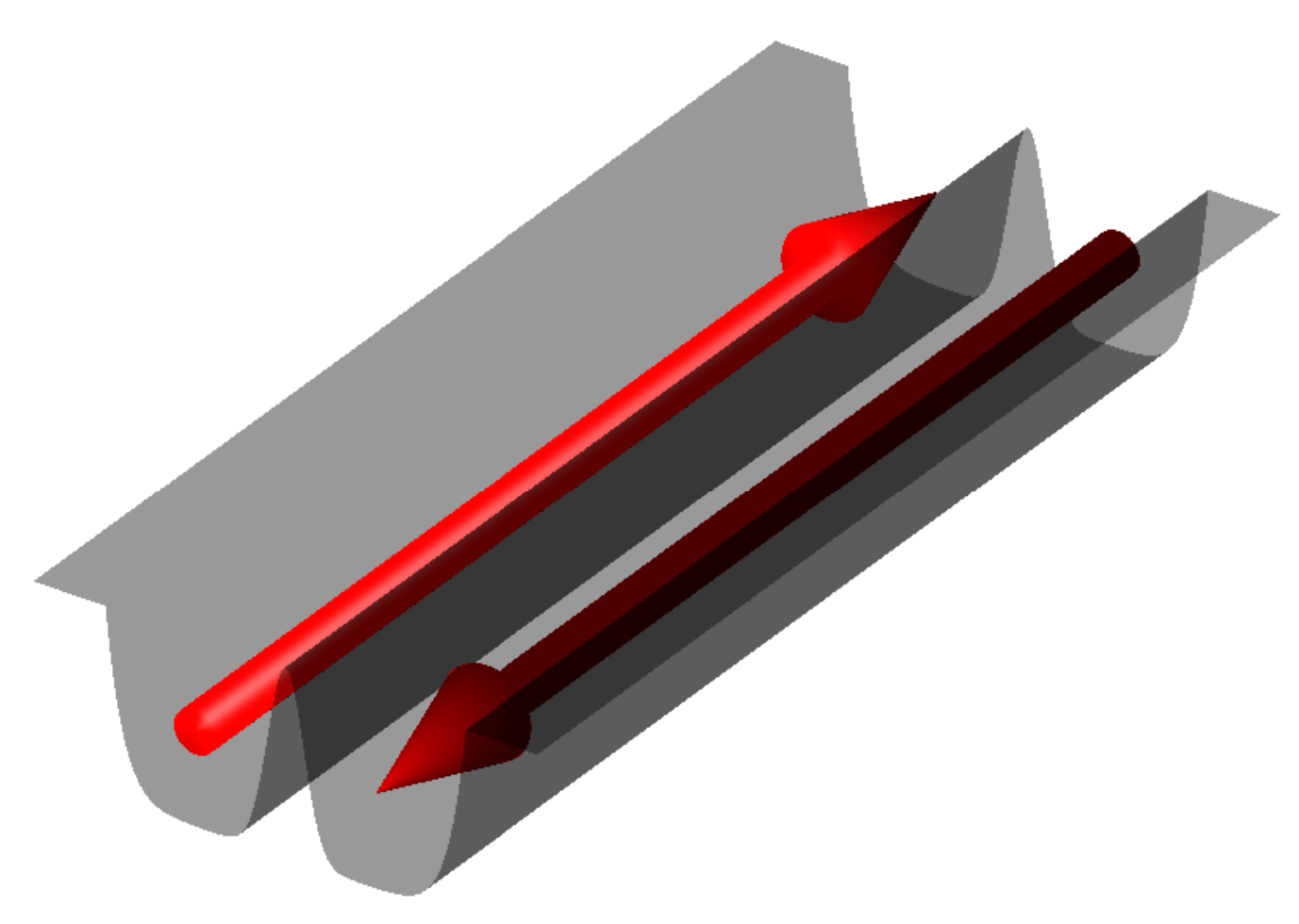}
\caption{A schematic of the initial configuration.  An atomic superfluid confined to a channel is divided by a central barrier; the superfluid on either side flows in opposite directions.  The central barrier is then lowered in time to create a region of high shear.}
\label{fig1}
\end{figure}
We choose a potential $V$ so as to create an overall channel aligned along $x$ which is separated into two sub-channels by a central barrier, as depicted in Fig.~\ref{fig1}.  The potential we take is uniform along $x$, and along $y$ it is a combination of a box potential and a central Gaussian potential, 
\begin{equation}
V(x,y,t) = V_B  \, \mathcal{H} \left(|y|-\frac{L}{2}\right)+V_G(t) \exp\left(-\frac{y^2}{\sigma^2} \right),
\end{equation}
where $\mathcal{H}$ denotes the Heaviside function. The box is taken to be $L=60$ wide and the potential walls are sufficiently high $V_B \gg 1$ to be effectively infinite.  Such box potentials can be realized experimentally using appropriately-shaped optical or electromagnetic fields \cite{boxes}.  The Gaussian potential is taken to have width $\sigma = 1.28$ and time-dependent amplitude, $V_G(t)$.  Such potentials can be created using focussed laser beams, with the amplitude controlled through the laser intensity. Initially $V_G=5$ such that the superfluid in each channel is separate from the other.

After numerically obtaining the condensate ground state (by imaginary-time propagation of the GPE \cite{Barenghi}), we impose a linearly-decreasing phase profile along $x$ in the $y>0$ sub-channel with total phase winding number $\mathcal{W}$.  This induces a uniform flow in the negative $x$ direction with speed $\pi \mathcal{W}/D_x$.  Similarly, we impose an equal and opposite flow in the $y<0$ sub-channel.  Note that the equal and opposite flow arrangement  is simply taken for convenience: our findings hold for any relative streamwise flow between the two sub-channels. 

If the potential barrier is maintained, the two fluids undergo persistent flow.  However, we choose to ramp the barrier down with time so as to merge the counter-propagating fluids and create a narrow region of large shear flow.   We ramp the barrier down according to the function $V_G(t)=\max(0,5-0.1t)$, although our findings are robust to changing the rate of this ramp-down.

\section*{Results}
 \begin{figure*}
\centering%
\includegraphics[width=0.245\linewidth]{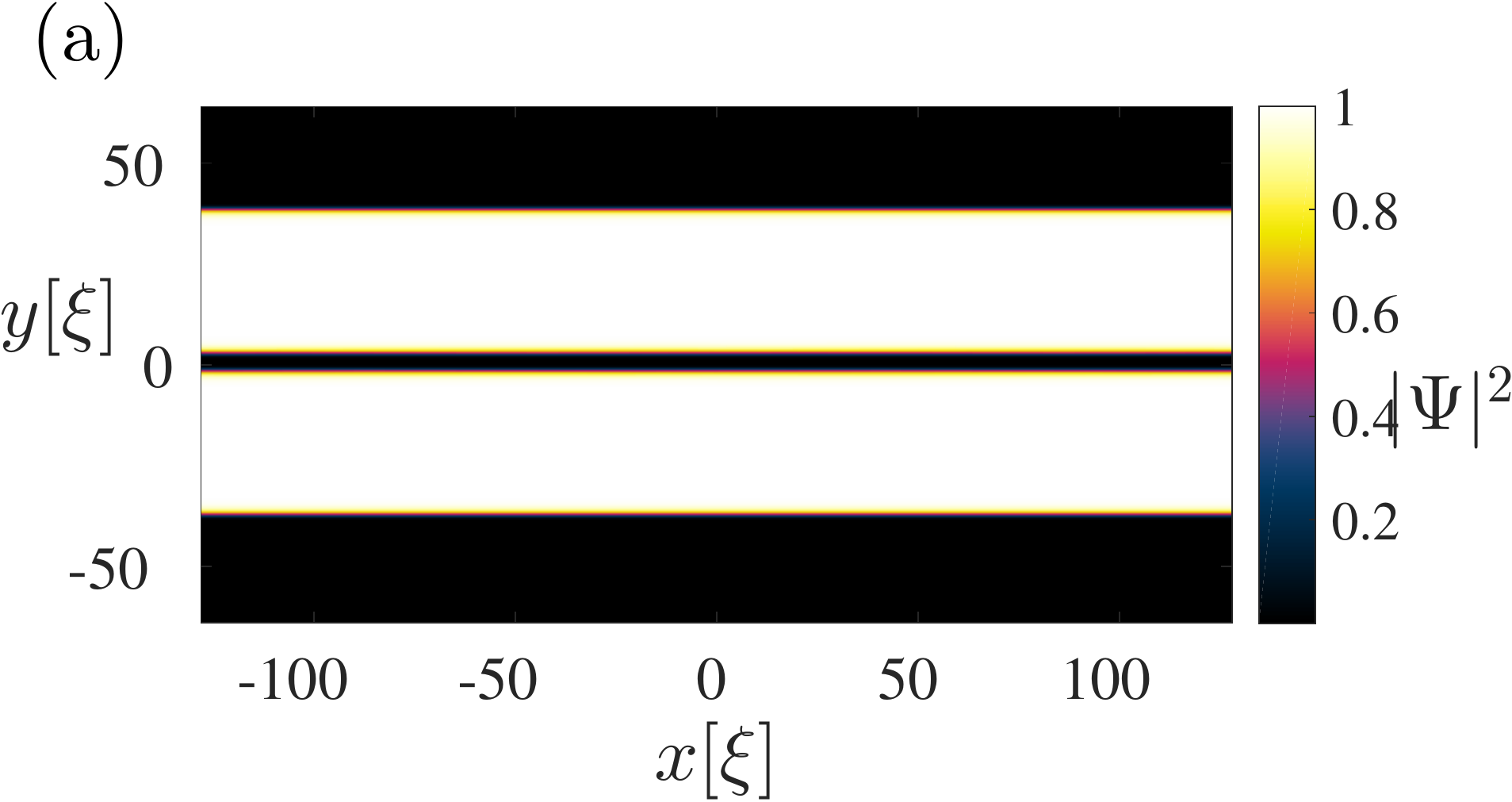}\hfill
\includegraphics[width=0.245\linewidth]{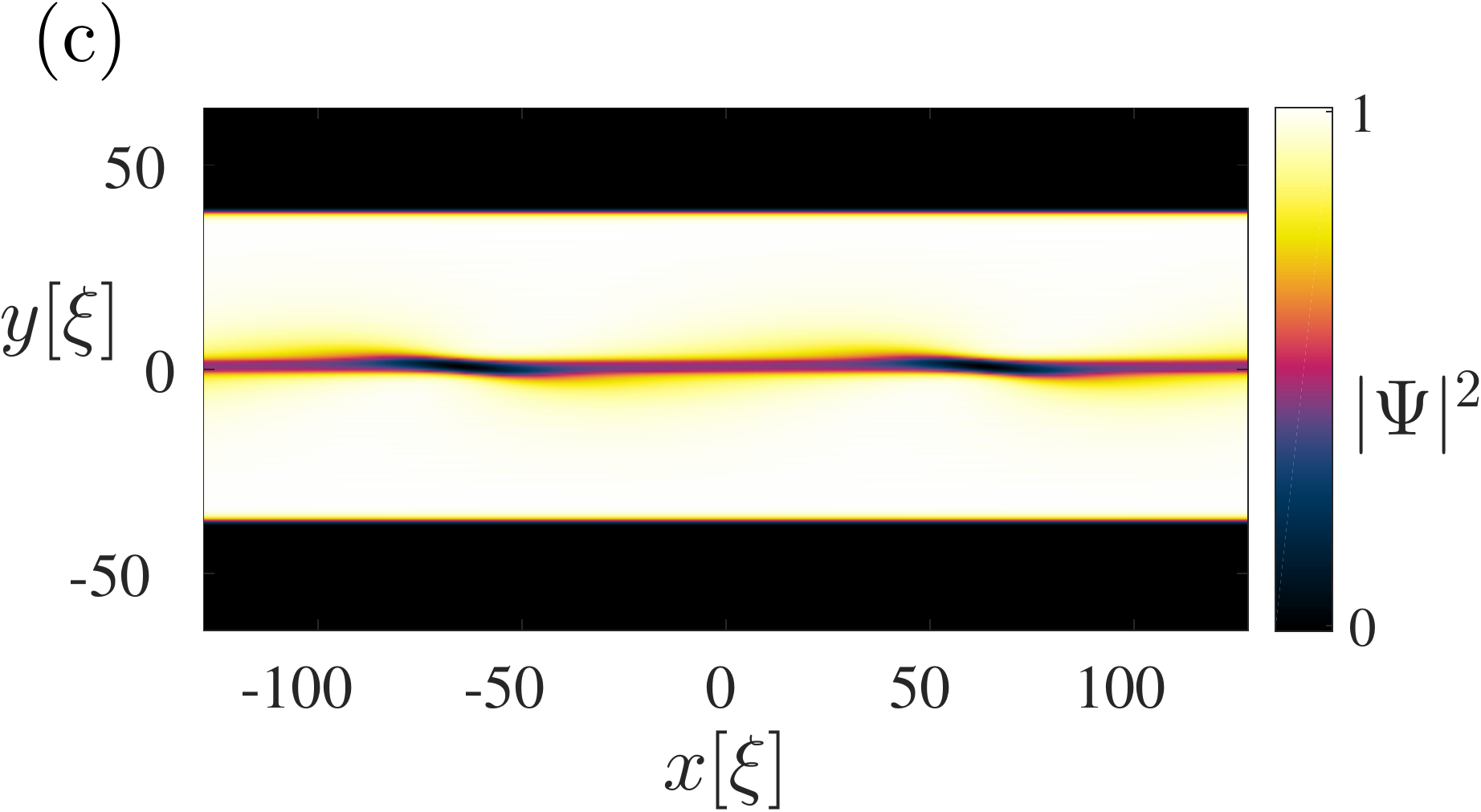}\hfill
\includegraphics[width=0.245\linewidth]{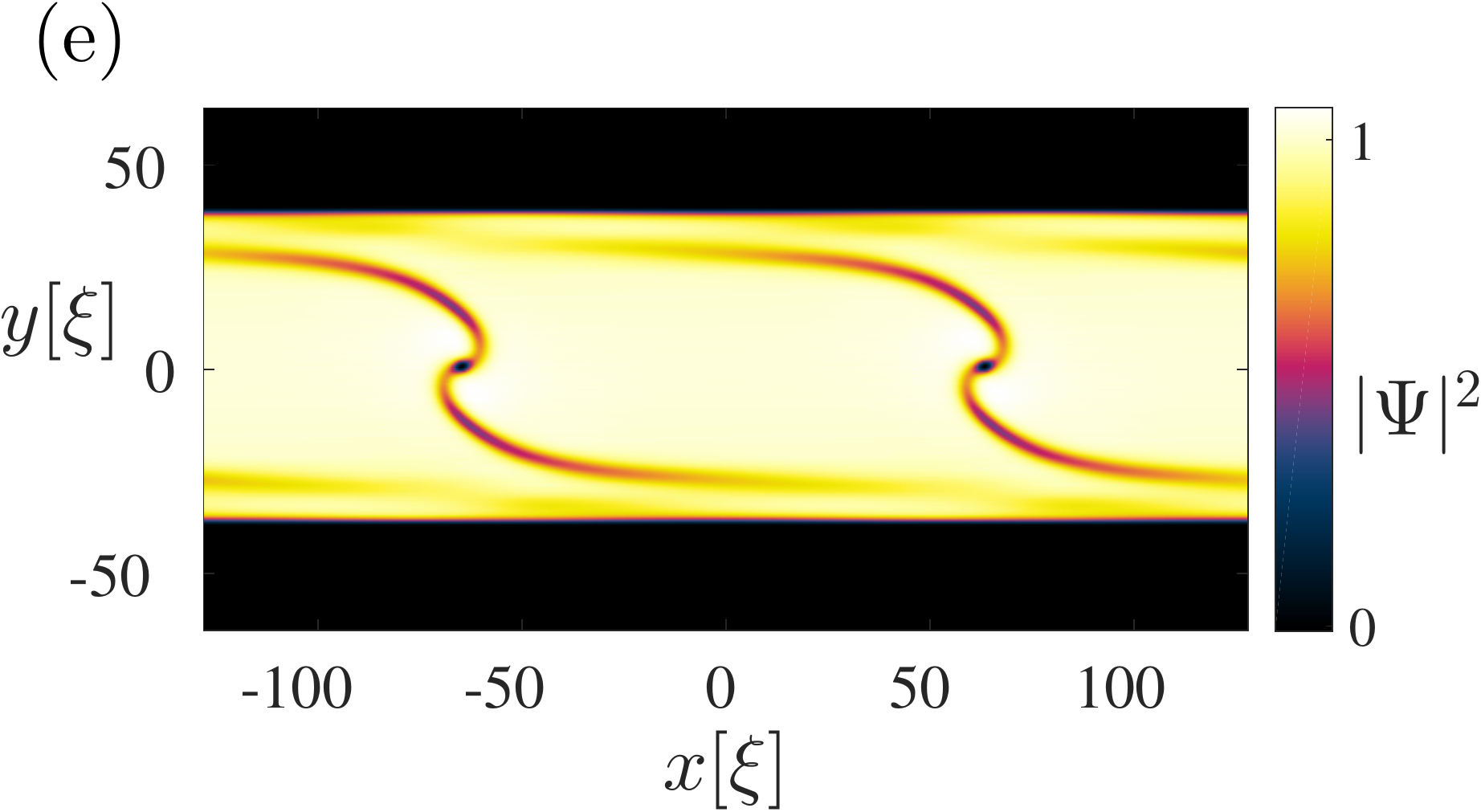}\hfill
\includegraphics[width=0.245\linewidth]{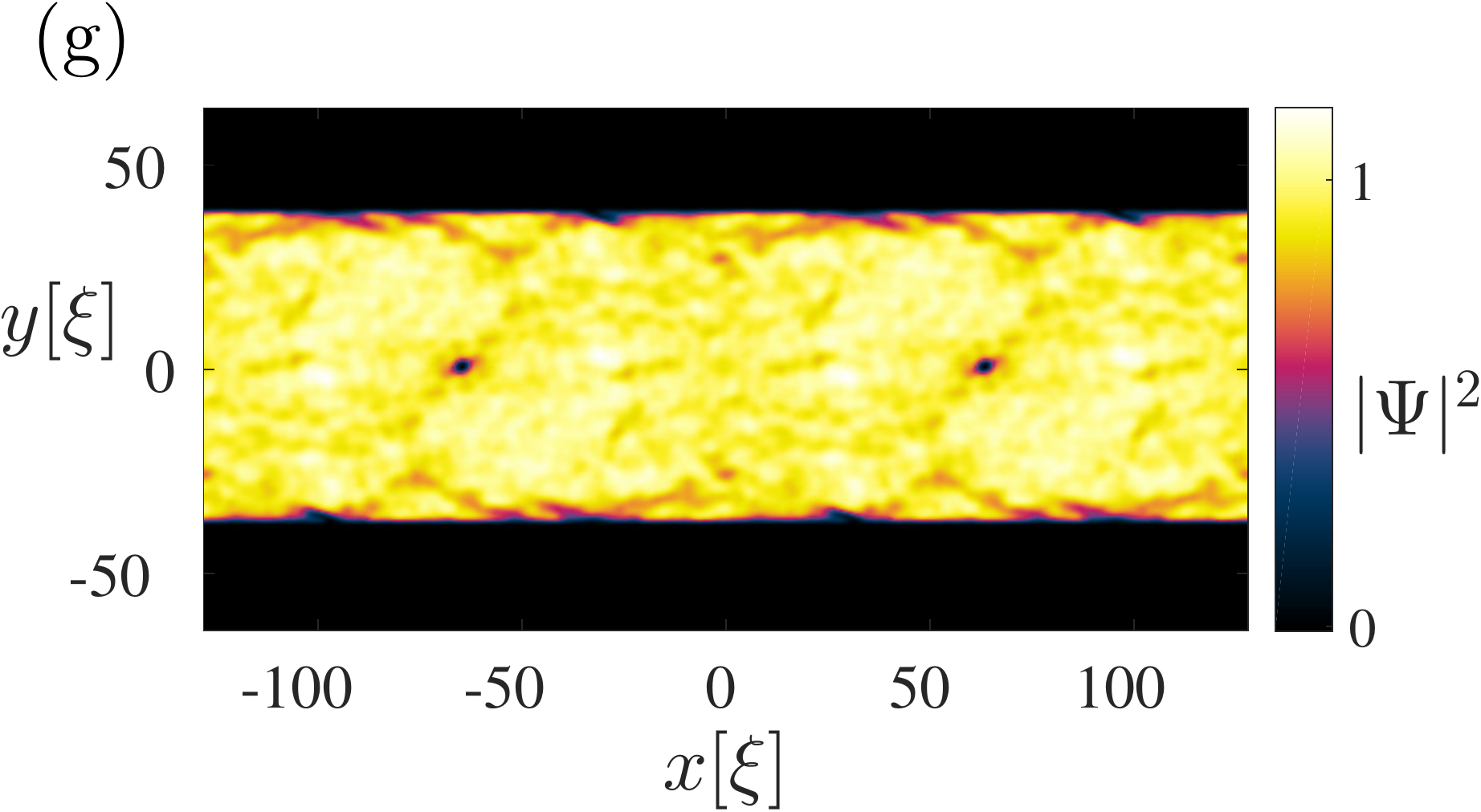}\\

\includegraphics[width=0.245\linewidth]{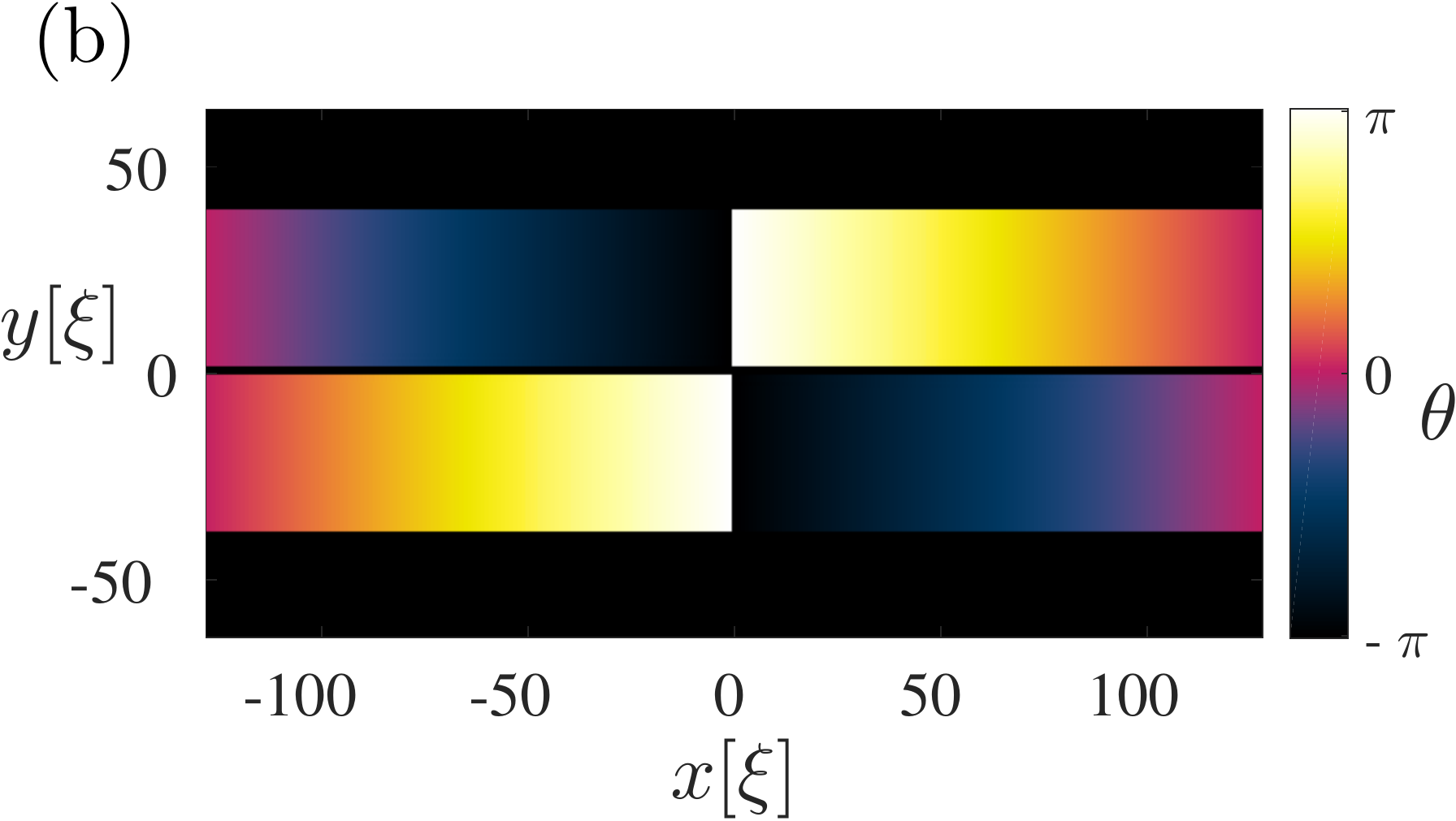}\hfill
\includegraphics[width=0.245\linewidth]{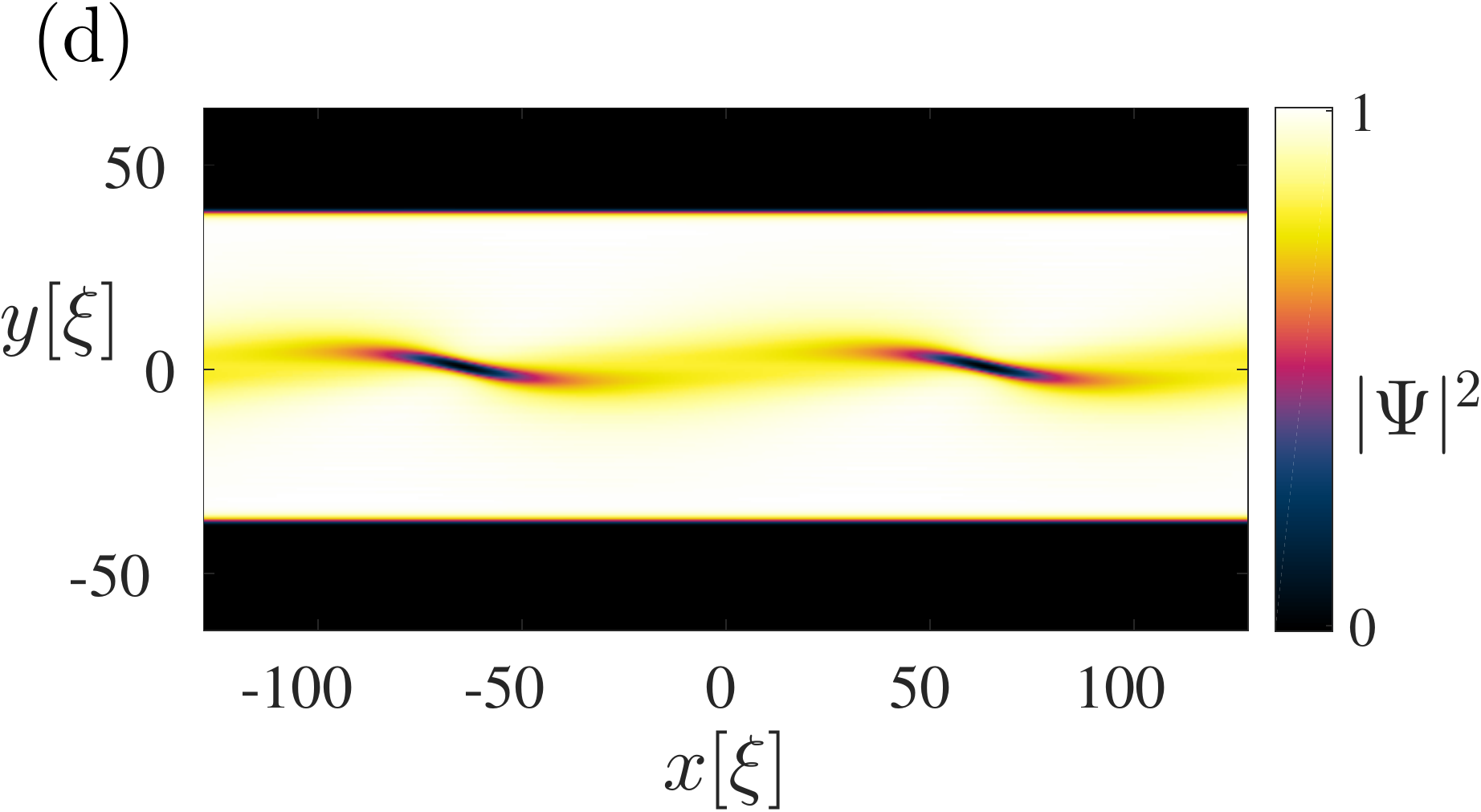}\hfill
\includegraphics[width=0.245\linewidth]{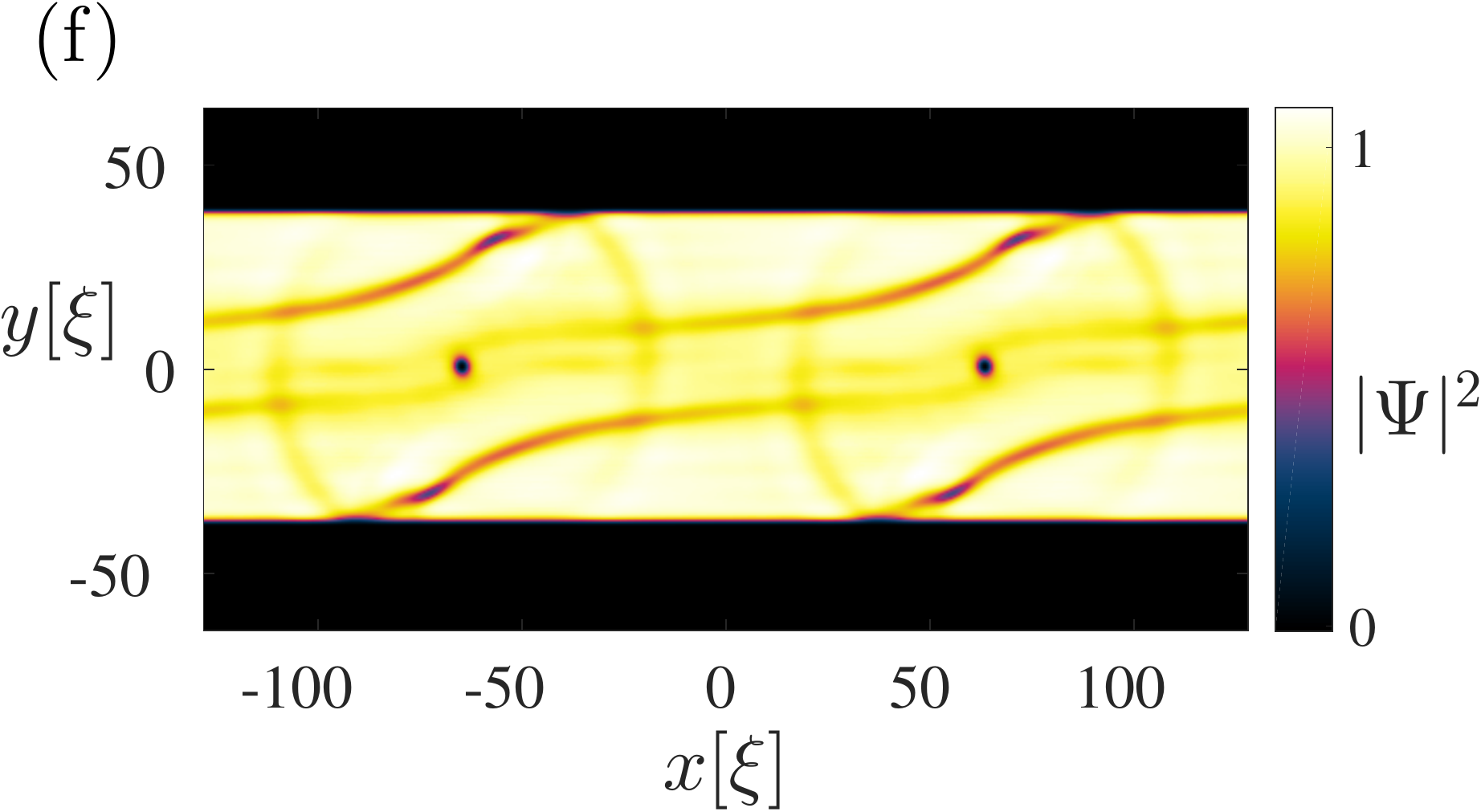}\hfill
\includegraphics[width=0.245\linewidth]{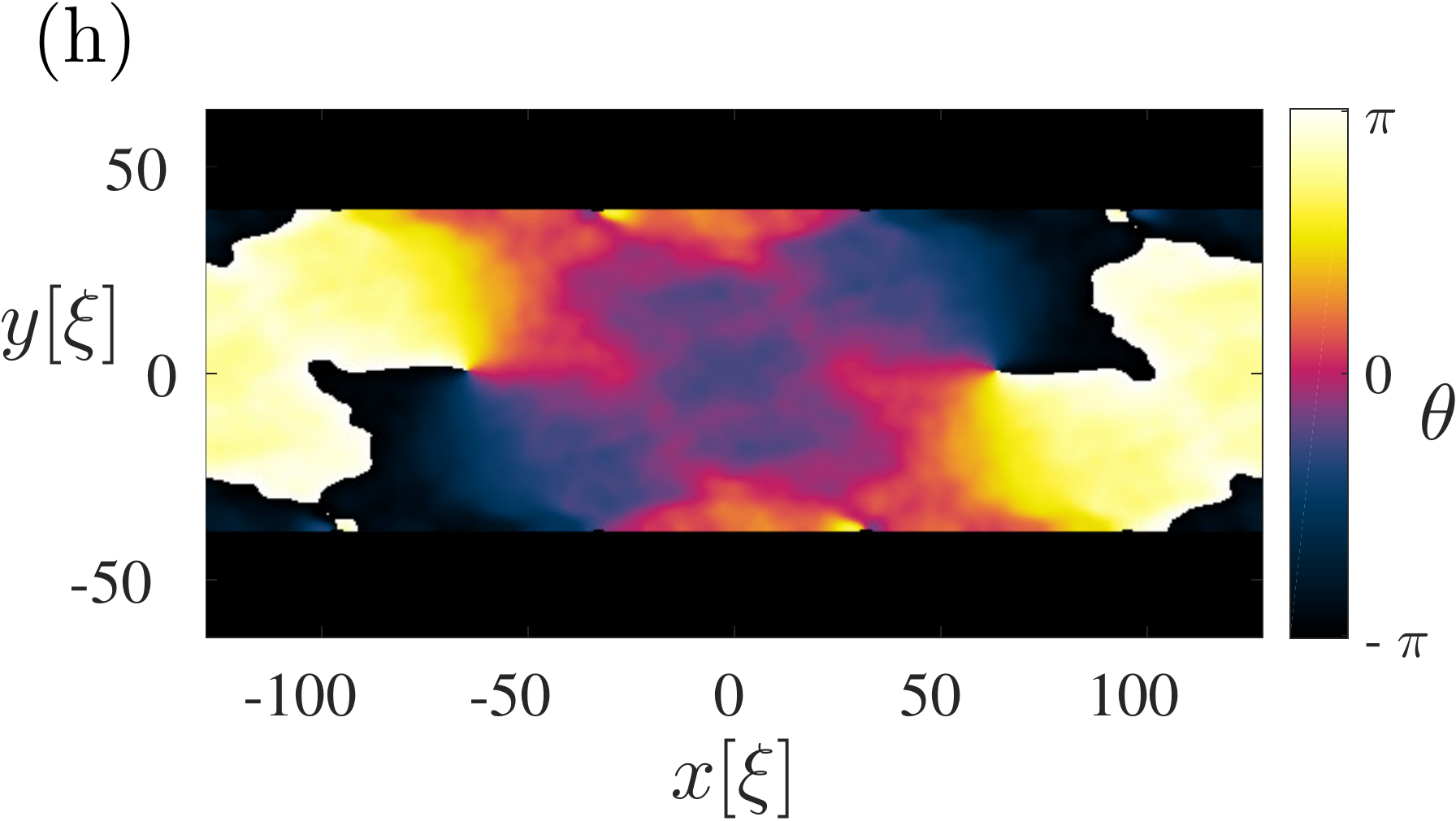}\\
\caption{Snapshots of the evolution of a simulation with winding number $\mathcal{W}=1$, in the absence of noise. Panels (a) and (b) show the density, $|\Psi|^2$, and phase, $\theta$ (for clarity only plotted where $|\Psi|^2>0.01$), at $t=0$. Panels (c)--(f) show the evolution of the density at times $45<t<150$ as the central barrier is lowered and two topological defects emerge. A final steady state density (g) and phase (h) are plotted at $t=600$.}
\label{fig2}
\end{figure*}
\begin{figure}[t]
\centering%
\includegraphics[width=0.8\linewidth]{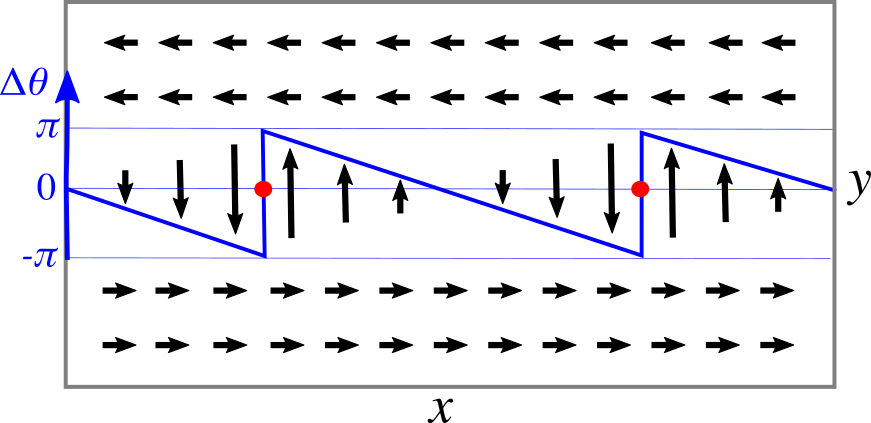}
\caption{Schematic of the formation of circulating flow at points (red dots) along the interface between the two flows.  The flow velocity is illustrated by black arrows and the phase difference across the interface $\Delta \theta$ by the blue line.}
\label{fig3}
\end{figure}

Figure \ref{fig2} shows the results for a winding number $\mathcal{W}=1$.  As the barrier drops the two persistent currents come into contact, exciting strong density perturbations. We see the formation of two topological defects, quantised vortices with the same sign of circulation, which remain in a stable configuration throughout the rest of the simulation. We have carried out simulations with up to $\mathcal{W}=30$ and all simulations result in the same end-state: a line of quantised vortices with some background phonon excitations.  Iit is clear that the number of vortices produced is simply $2\mathcal{W}$.

The explanation for the formation of $2 \mathcal{W}$ like-signed vortices is straightforward and illustrated in Fig. \ref{fig3}.  Across the $y=0$ interface there exists a discontinuity in the condensate phase $\theta$.  This phase difference, defined as $\Delta \theta=\theta(x,y=0^+)-\theta(x,y=0^-)$ (mod $2\pi$), has a saw-tooth profile along the channel (due to the wrapping of the phase between $-\pi$ and $\pi$).   Now recall that the fluid velocity is proportional to the gradient of the phase, and hence this gives rise to a saw-tooth-profile velocity component along $x$.  At points along the inteface where the phase jumps by $2\pi$ this velocity component discontinuous switches direction.  There are exactly $2\mathcal{W}$ such points along the interface.  When coupled with the imposed flows for $y>0$ and $y<0$, this gives rise to a circulating flow around these points on the interface, which hence immediately evolve into quantised vortices.  

\begin{figure*}
\centering%
\includegraphics[width=0.32\linewidth]{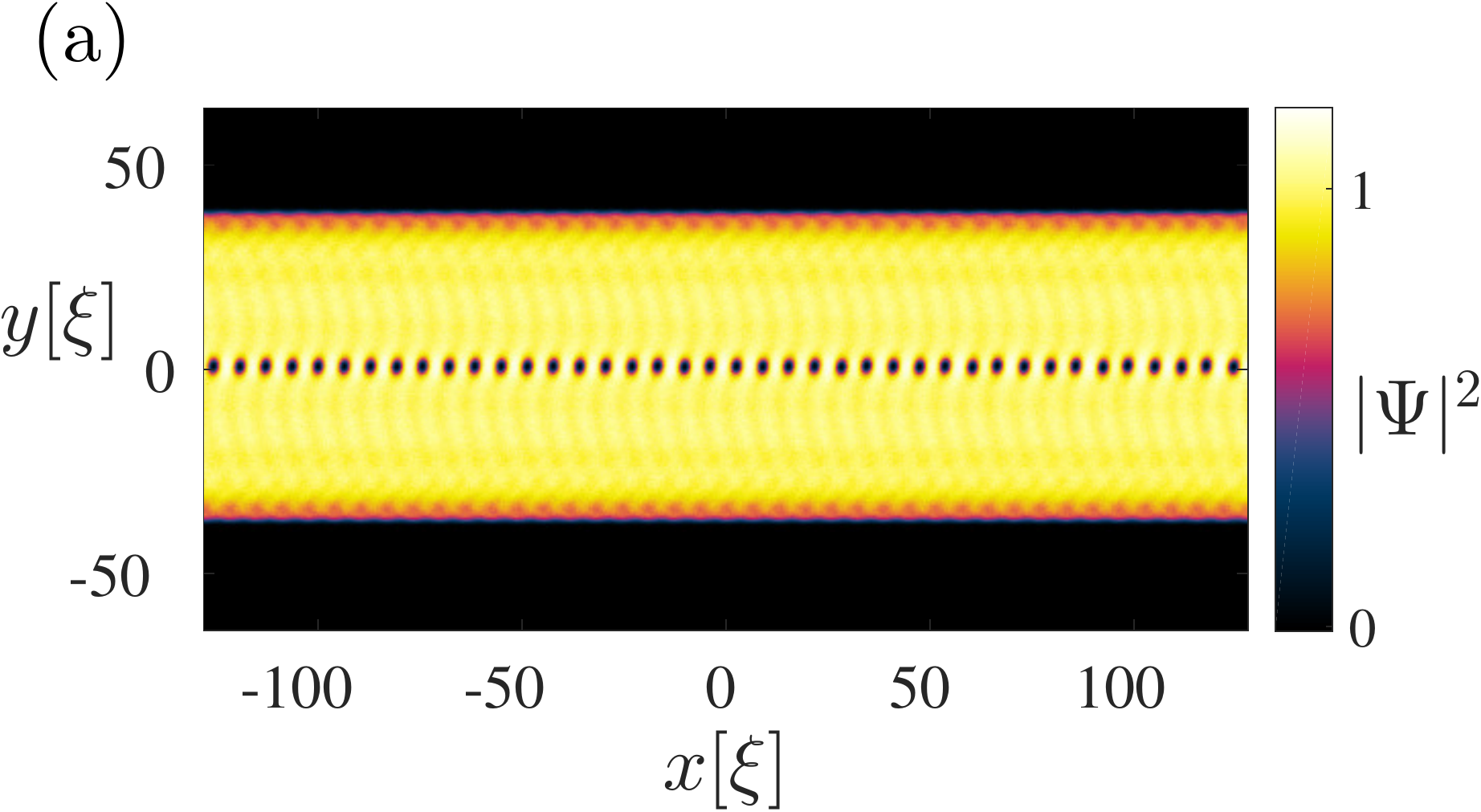}
\includegraphics[width=0.32\linewidth]{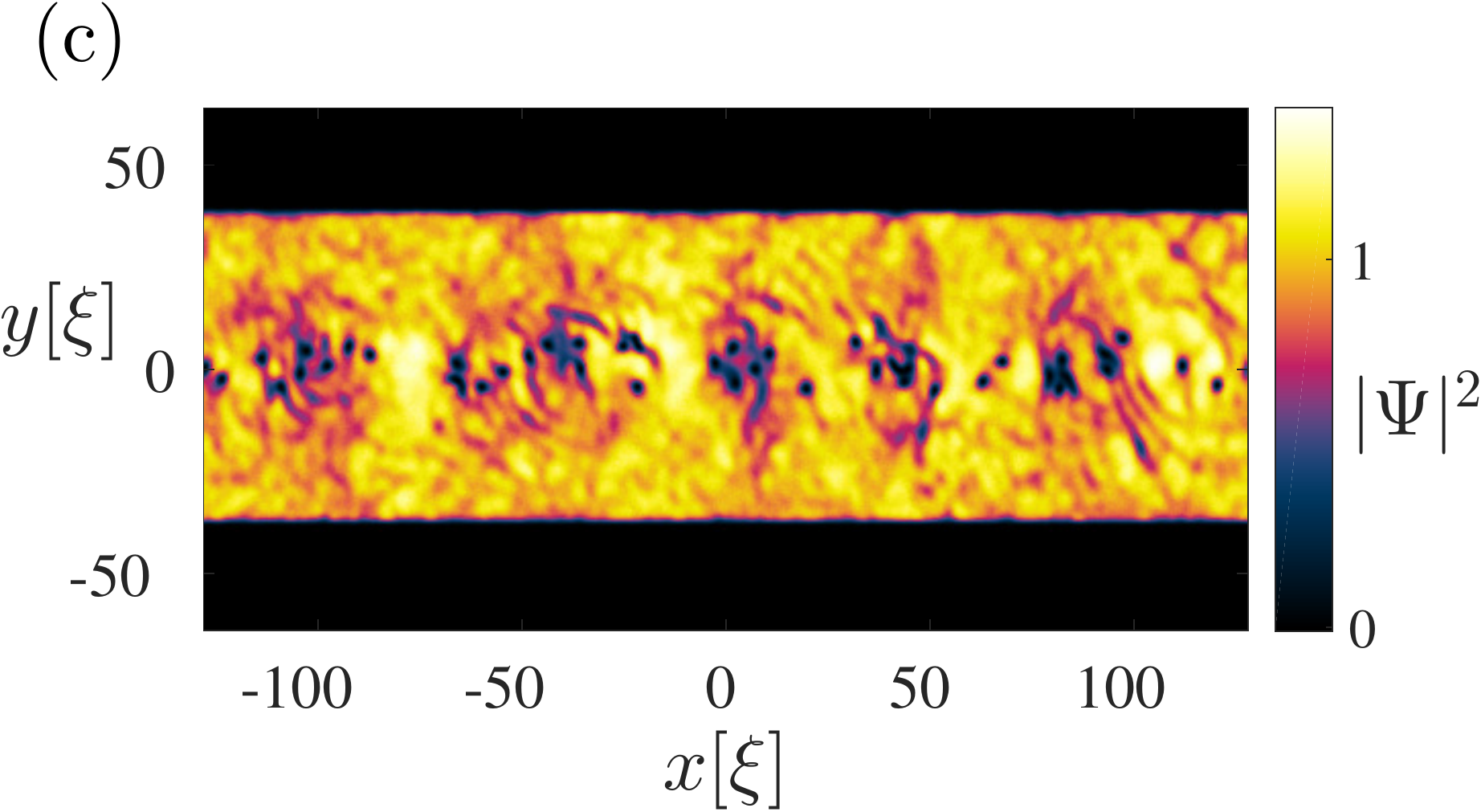}
\includegraphics[width=0.32\linewidth]{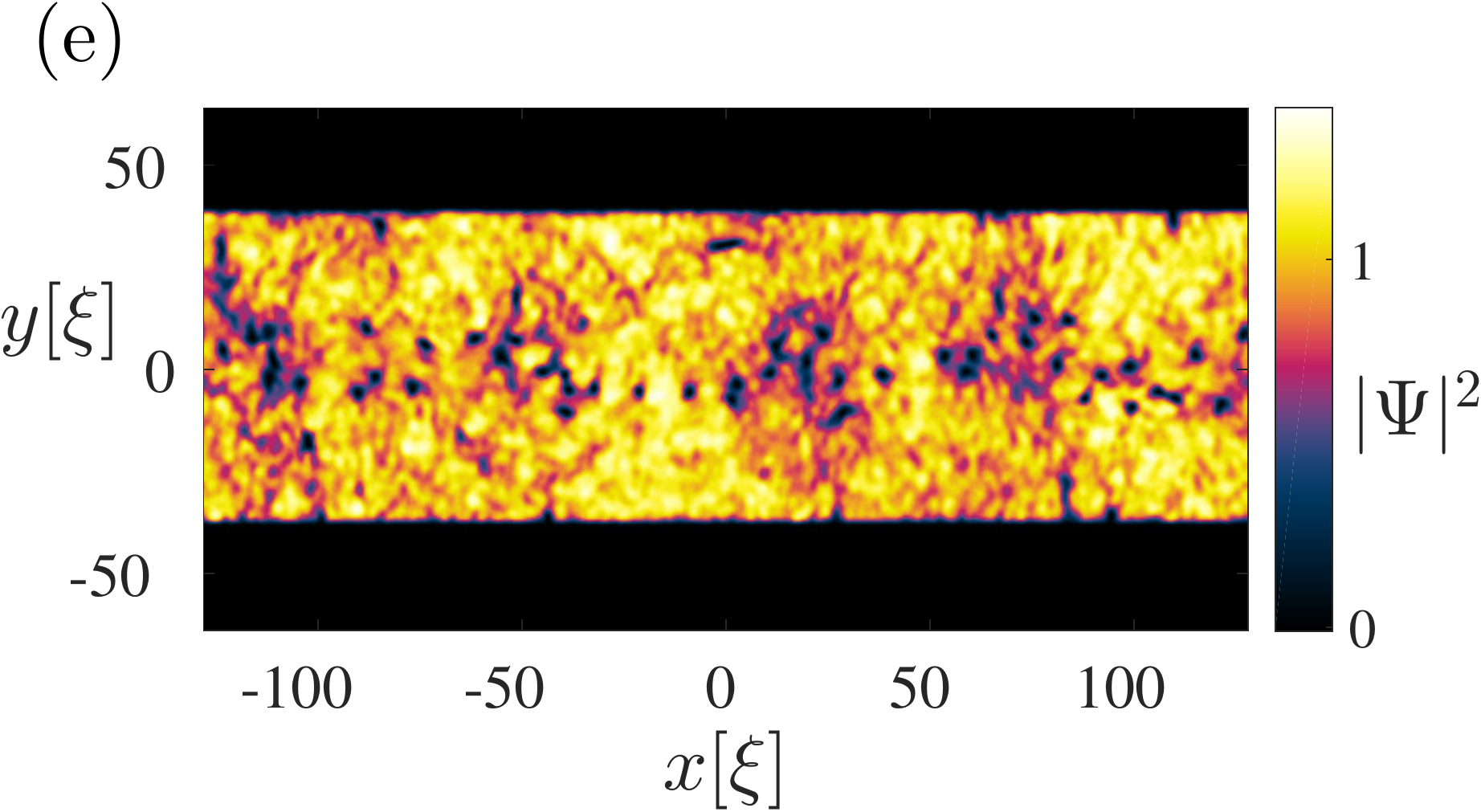}\\

\includegraphics[width=0.32\linewidth]{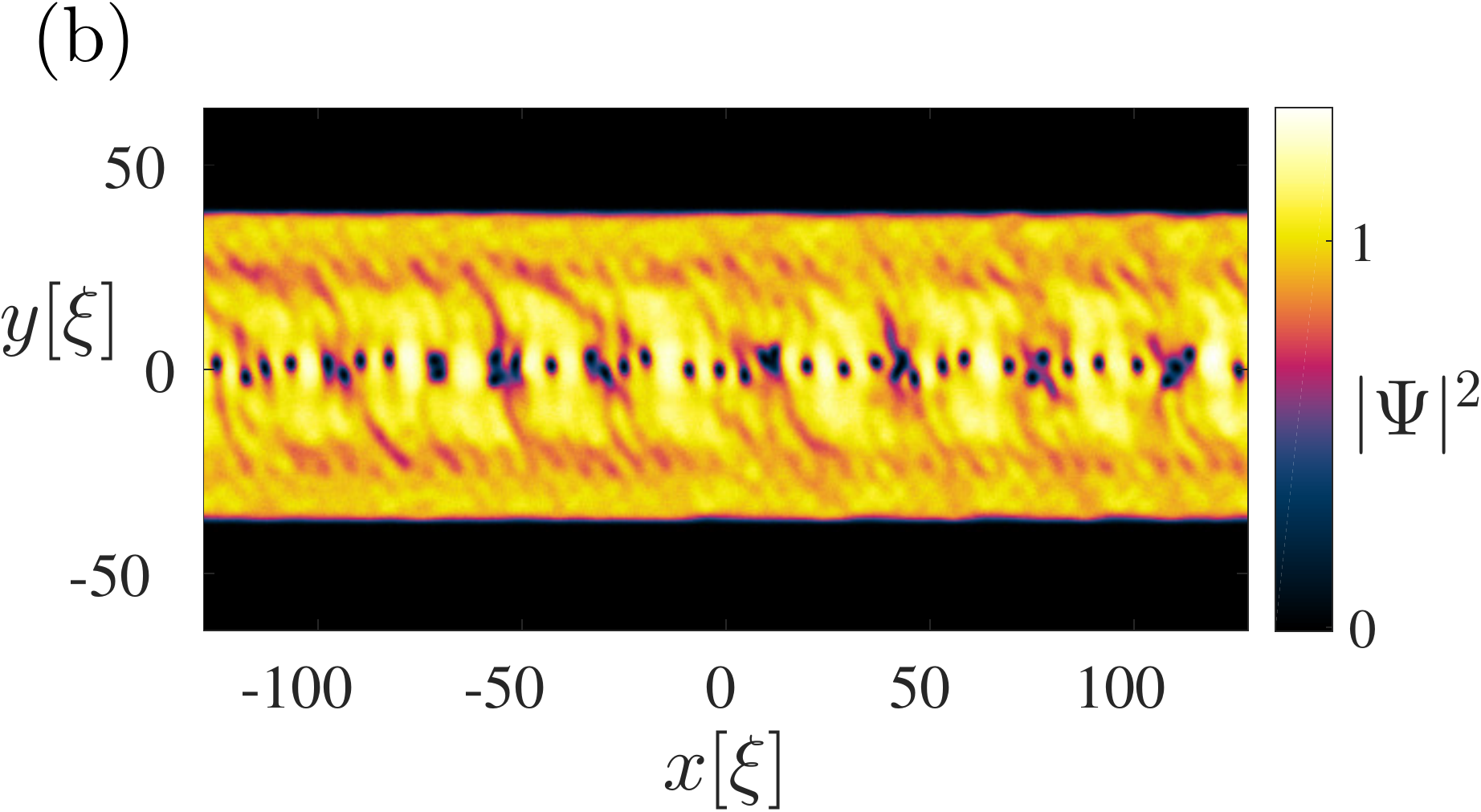}
\includegraphics[width=0.32\linewidth]{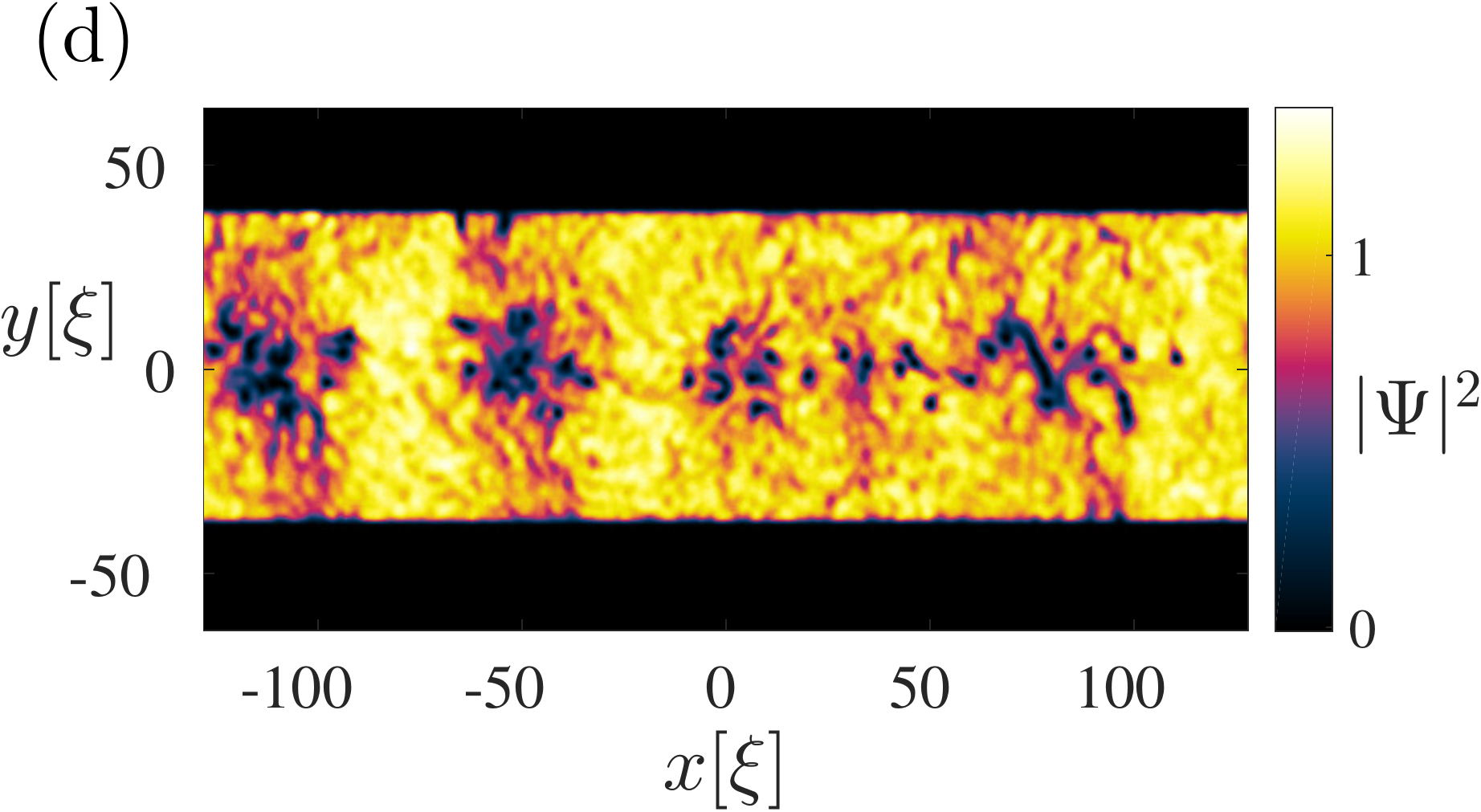}
\includegraphics[width=0.32\linewidth]{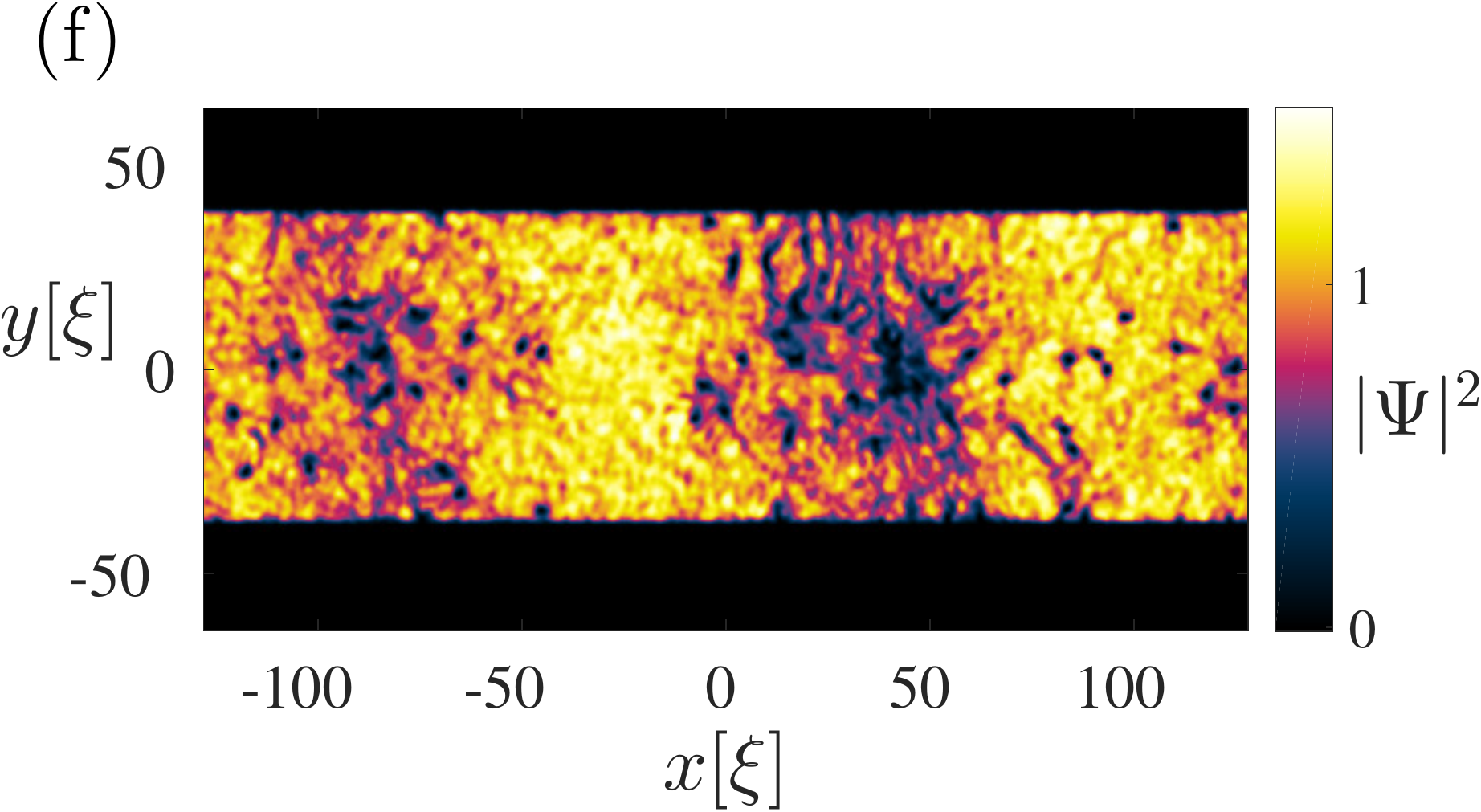}

\caption{Snapshots of the evolution of the density with $\mathcal{W}=20$, with a small amount of noise added to the initial condition, for (a) $t=100$; (b) $t=200$; (c) $t=300$; (d) $t=400$; (e) $t=500$; (f) $t=700$. Note the formation of a quantum vortex sheet which `rolls-up' via a Kelvin-Helmholtz instability.}
\label{fig4}
\end{figure*}

What is produced is the quantum analogue of a classical vortex sheet. Whereas a classical vortex sheet is a continuous curve along which the fluid vorticity is non-zero, the quantisation of vorticity in a superfluid prevents this and instead supports a line of quantised vortices.  It is interesting to note that in studies of classical vorticity, vortex sheets are often computed as collections of point vortices along a curve \cite{Krasny}; thus the quantum vortex sheet is a direct realization of this mathematical abstraction.  We also note that vortex sheets have been predicted in two-component BECs \cite{Kasamatsu}.

Subject to perturbations we would expect the quantum vortex sheet to roll-up via a KH instability. In our simulations we find without the presence of an internal or external perturbation the quantum vortex sheet remains stable, at least for as long a time period as it is feasible to integrate for. However the addition of a small amount of white noise (whose magnitude is less than 1\% of the background wavefunction) to our initial configuration is sufficient to realise a KH instability in a single component superfluid. In a real system, such noise will be present from a variety of sources, including thermal, quantum and mechanical effects.

\begin{figure}
\begin{center}
\includegraphics[width=0.98\columnwidth]{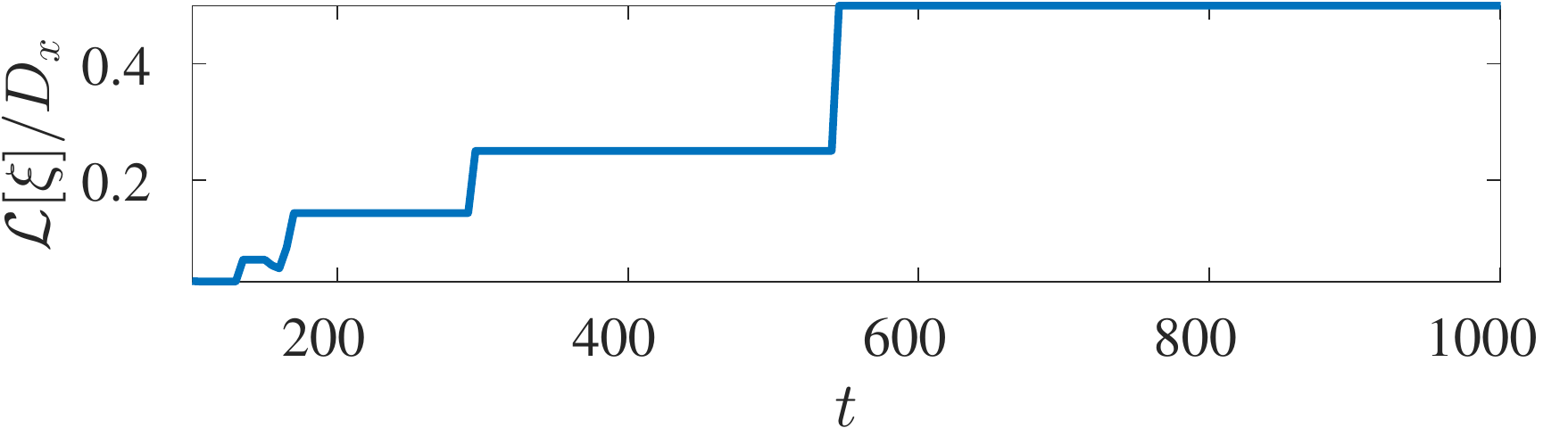} 
\caption{The evolution of the spatial extent of vortex cluster, $\mathcal{L}$, in time.}
\label{fig5}
\end{center}
\end{figure}

Figure \ref{fig4} shows the evolution of a simulation with $\mathcal{W}=20$, with a small amount of noise added to the initial condition.  As expected from above, the interface rapidly evolves into a quantum vortex sheet of 40 like-sign vortices.   The vortex line then visibly destabilises.  The vortices first tend to bunch up into small clusters of 2-4 vortices, which co-rotate.  Over time, the clusters merge with neighbouring clusters, forming progressively bigger clusters.  This process is the quantum analog of the progressive roll-up of a classical vortex sheet, in others words, the KH instability.  It is worth noting that the clusters of many like-signed vortices act to mimic classical patches of vorticity.

To monitor the effective cluster size we first integrate the atomic density in the transverse direction, defining
\begin{equation}\label{eq:navg}
\rho (x,t)=\langle n(x,y,t) \rangle_y=\frac{1}{2 D_y} \int_{-D_y}^{D_y} n(x,y,t) dy,
\end{equation}
where we interpret $\rho$ as a course-grained density field. Denoting $P(k,t)$ as the Fourier-transform of $\rho$, then the typical spatial extent of a cluster can be estimated as $\mathcal{L}(t)=D_x/2 |\hat{k}(t)|$, where
\[
\hat{k}(t)=\argmax_{|k|>0} P(k,t).
\]
Figure \ref{fig5} shows the evolution of $\mathcal{L}$ in time.  The step-wise increase of the cluster size $\mathcal{L}$ is consistent with the progressive merger of smaller clusters into larger ones of approximately double the size.  This process ceases when the clusters become comparable in size to half the length of the channel. Note how the merging becomes slower as the cluster size increases.

\begin{figure}
\begin{center}
\includegraphics[width=0.4\textwidth]{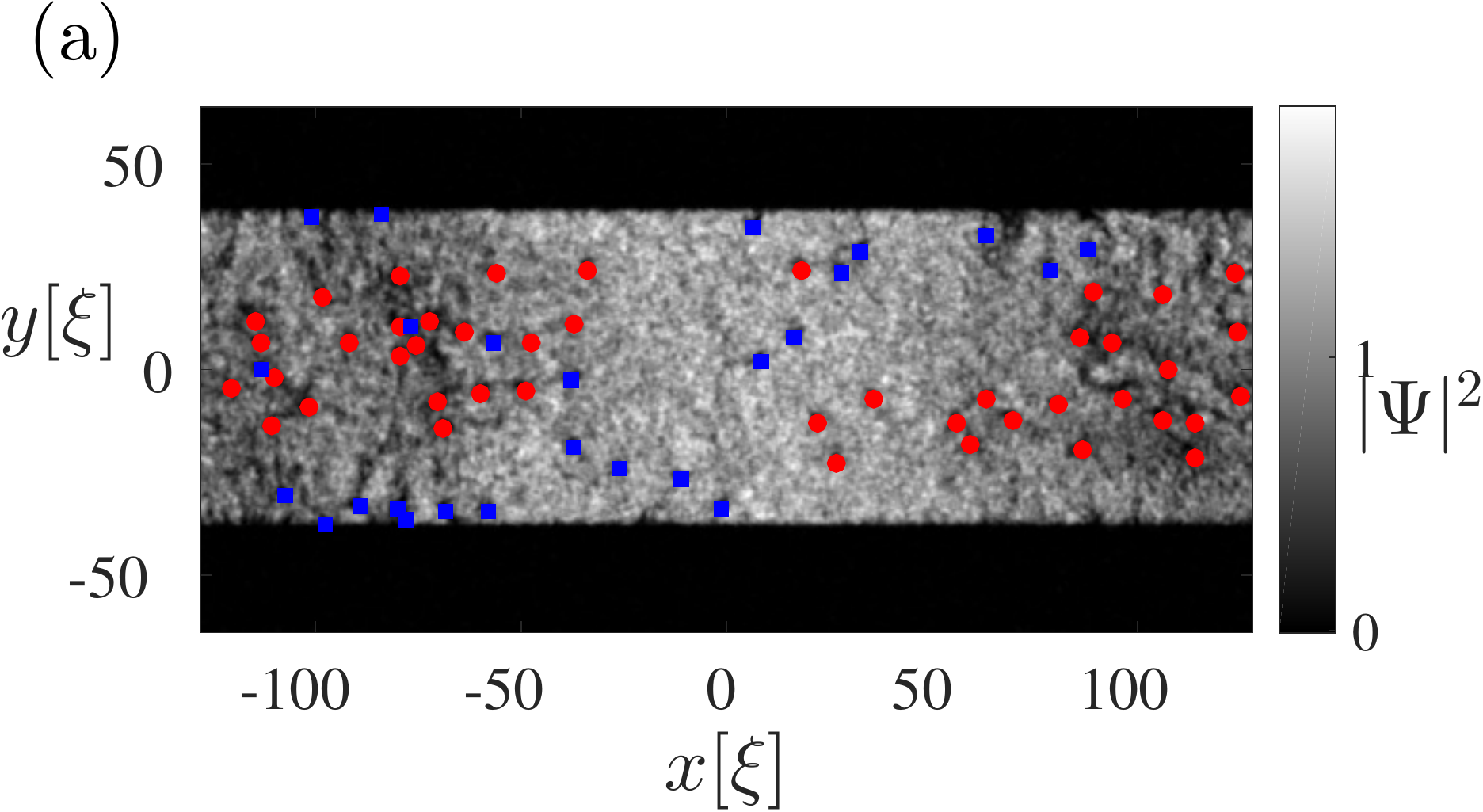} \\
\includegraphics[width=0.4\textwidth]{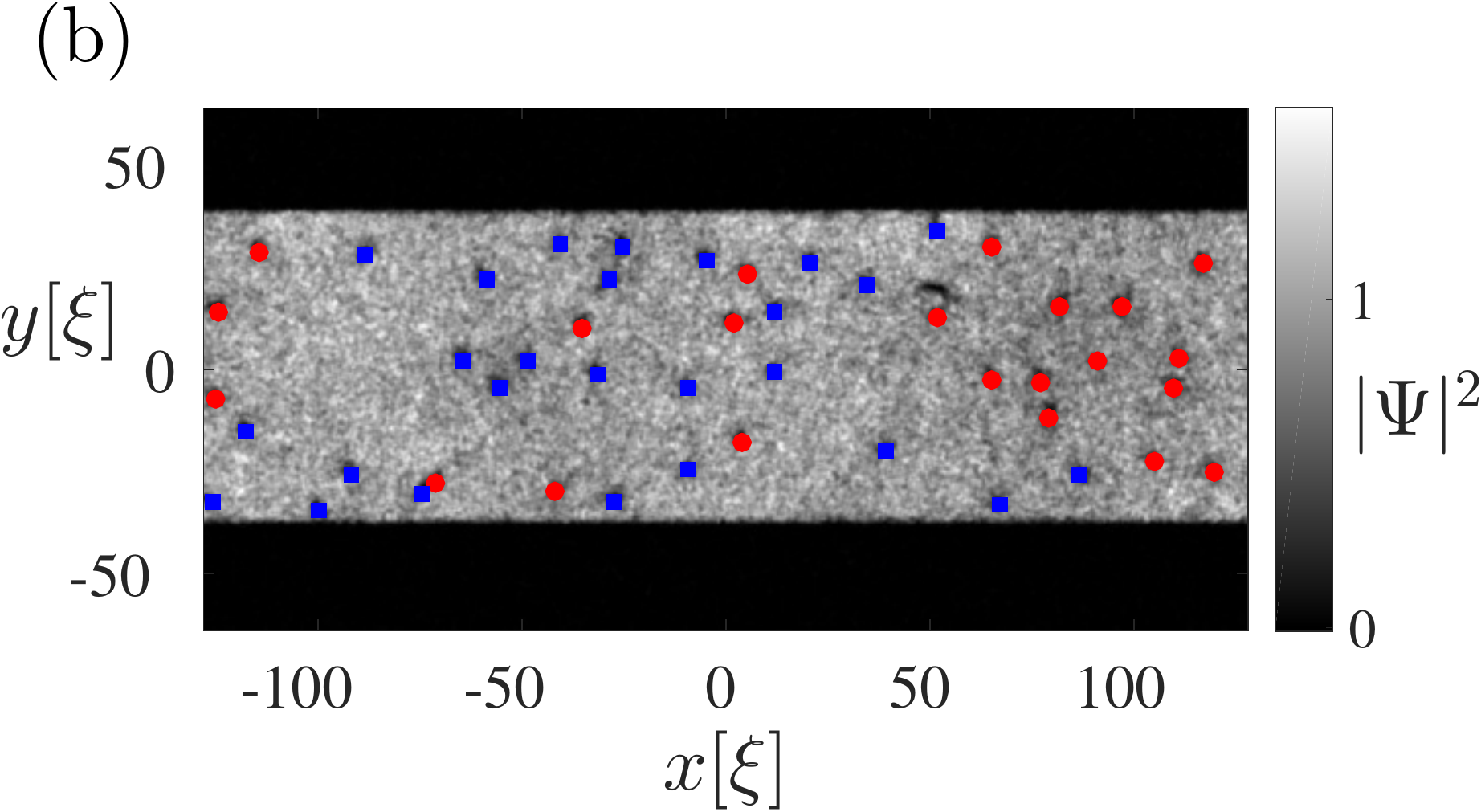} \\
\caption{Snapshots of the late time evolution of the system with $\mathcal{W}=20$, at times (a) $t=1800$ and (b) $t=5000$. Red circles and blue squares mark the location of vortices with positive and negative circulation respectively.}
\label{fig6}
\end{center}
\end{figure}

Up until this point the vortices are dominately of the same circulation, which is the circulation of the vortices in the initial quantum vortex sheet.  The KH instability is interrupted as the vortices try to form a single large cluster. Angular momentum is not conserved in our system due to the presence of the external potential. This exerts a torque on the gas, which manifests itself through the appearance of negatively signed vortices which are created at the edge of the condensate and penetrate into the bulk. Over time this collection of vortices of positive and negative sign evolve into a quasi-steady-state composed of clusters of like-sign vortices, see Fig.~\ref{fig6}.  These clusters are consistent with negative-temperature Onsager vortex clusters, which were originally predicted to be the preferred state of high-energy two-dimensional turbulence of point vortices \cite{Onsager}.  More recently, these states have been shown to arise in atomic BECs \cite{Simula2014,Billam2014}, including recent experimental observations \cite{Johnstone2018,Gauthier2018}.

\begin{figure}
(a) ~~~~~~~~~~~~~~~~~~~~~~~~~~~~~~~~~~~~~~~~~~~~~~~~~~~~~~~~~~~~~~~~~~~~~~~~~~~~~~
\\
\includegraphics[width=0.95\linewidth]{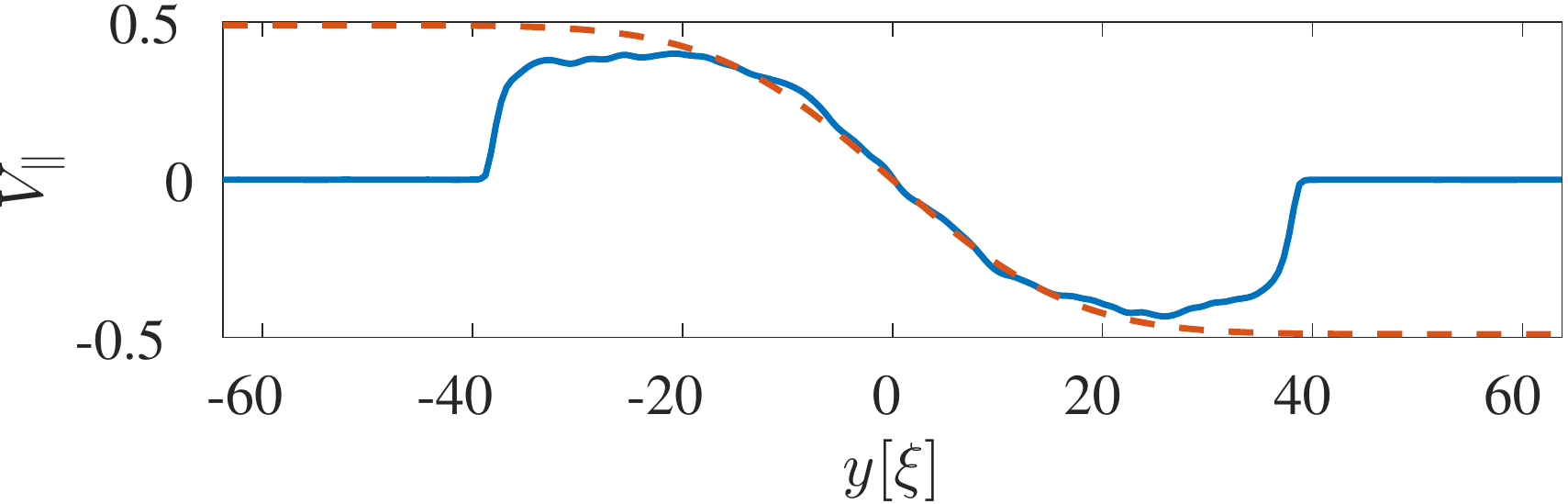}\\
(b) ~~~~~~~~~~~~~~~~~~~~~~~~~~~~~~~~~~~~~~~~~~~~~~~~~~~~~~~~~~~~~~~~~~~~~~~~~~~~~~
\\
\includegraphics[width=0.95\linewidth]{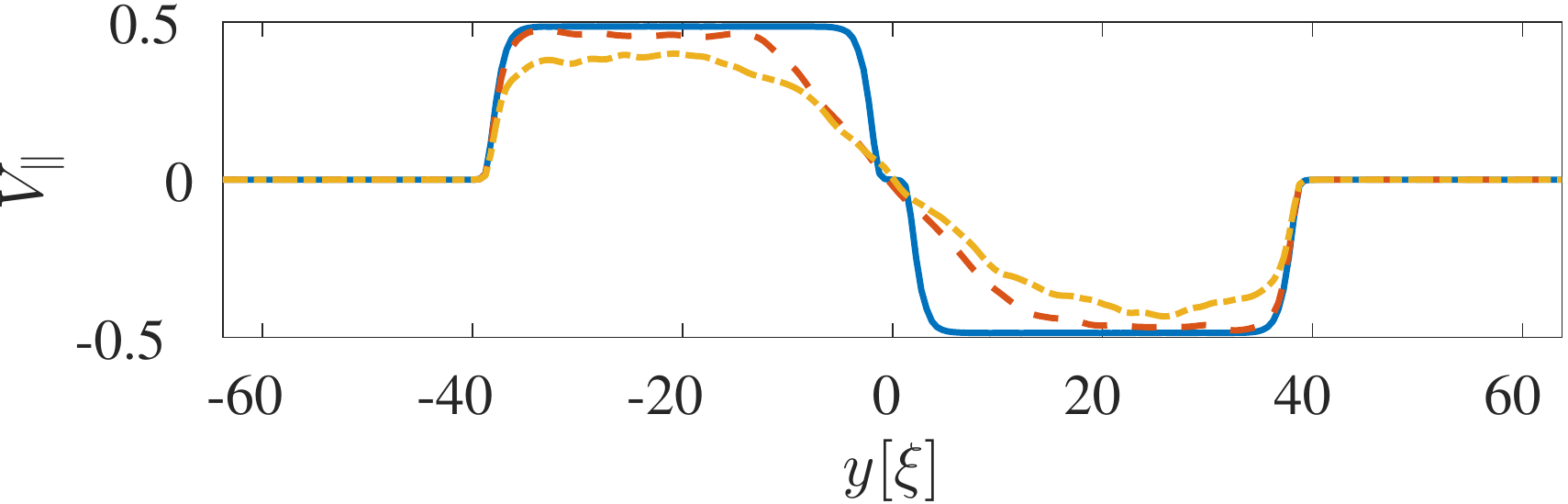}
\caption{The evolution of the course-grained streamwise component of the momentum along the channel,  $V_\parallel=\langle m v_x\rangle_x$.
(a) The solid line shows the two counter-flowing streams at $t=0$; the dashed line shows the existence  of a superfluid shear flow at $t=500$; its subsequent evolution (due to effective viscosity) is shown at $t=1000$ via the dot-dashed line.
(b) The solid line plots $V_\parallel$ at $t=1000$ computed from the simulations, the dashed line displays the solution to the diffusion equation Eq.~(\ref{eq:diffsoln}), with $\nu'=0.1$.
}
\label{fig7}
\end{figure}

To further study how this flow mimics its classical counterpart, we next examine the coarse-grained momentum of the fluid, integrated along the channel. Indeed, we can readily compute the (dimensionless) momentum directly from the wavefunction via $n \mathbf{v}=i\left(\Psi \nabla \Psi^*-\Psi^* \nabla \Psi\right)$.  We integrate the streamwise component of the momentum along the channel, denoting this quantity $V_\parallel=\langle n v_x\rangle_x$, where $\langle \cdot \rangle_x$ represents averaging over the streamwise dimension (see Eq.~(\ref{eq:navg}) for the precise definition of the averaging represented by the angled brackets).
Figure \ref{fig7} (a) shows the evolution of this quantity, computed from three snapshots as the simulation progresses. Before lowering the central barrier $V_\parallel$ corresponds to a superposition of that in the two independent sub-channels, where each flow has $V_\parallel\approx \pm 0.5$.  However, once the barrier is lowered we see a smoothening of the transition, which is approximately linear in the vicinity of $y=0$.  This is akin to a classical viscous shear layer between two regions of fluid under relative motion.  While in the classical case, shear layers are supported by shear forces and viscosity, in the viscosity-free superfluid this analogous behaviour is generated by the collective action of the many quantised vortices.  Similarly, the superfluid analogous of a boundary layer was recently predicted in the form of the collective behaviour of many vortices close to the surface \cite{Stagg2017}.

This analogy to 1D classical shear flow provides a means to estimating the effective viscosity of the quantum fluid.  
Effective viscosity, $\nu'$, is a widely used concept in both experimental and theoretical studies of superfluid helium \cite{Walmsley,Stagg}, where it is commonly used to interpret the dissipation of (incompressible) kinetic energy. At extremely low temperatures where thermal dissipation mechanisms are negligible, dissipation arises from phonon emission when vortices accelerate (due to the influence of other vortices, density inhomogeneities or boundaries) \cite{vinen2001,barenghi2005} or reconnect/annihilate with each other \cite{zippy2001,stagg2015,baggaley2018}.  

In this 1D limit the Navier-Stokes equation for a classical viscous fluid reduces to a simple diffusion equation, and so we can estimate the effective viscosity, $\nu'$, by comparing our evolution of $V_\parallel$ to solutions of the 1D diffusion equation,
\begin{equation}\label{eq:diff}
\frac{\partial V_\parallel}{\partial t}=\nu' \frac{\partial^2 V_\parallel}{\partial y^2}.
\end{equation}
If we assume the initial form for $V_\parallel$ to be
\[
    V_\parallel(y,0) =\begin{cases}
        \pi\mathcal{W}/D_x&y<0\\
        -\pi\mathcal{W}/D_x&y>0
    \end{cases},
\]
then the solution to Eq.~(\ref{eq:diff}) is simply
\begin{equation}\label{eq:diffsoln}
V_\parallel=\frac{\pi\mathcal{W}}{D_x}\, \mathrm{erf}\left(\frac{y}{\sqrt{4 \nu' t}}\right).
\end{equation}
We obtain $\nu'$ by fitting this analytic solution to $V_\parallel(y)$ from our course-grained simulations, with the fit shown in Fig. \ref{fig7}(b). Hence we estimate $\nu' \approx 0.1$. Given our non-dimensional quantum of circulation is $\kappa=2\pi$ we estimate $\nu'/\kappa \approx 0.015$, which we can compare to values from the literature.

Before proceeding it is important to note that the estimates of $\nu'$ to date come from three-dimensional studies of superfluid helium, which is very different from the fluid system in this study. With that caveat in mind, the most complete compilation of $\nu'$ to date is found in  \cite{Walmsley}, who show that in the limit of zero temperature $\nu'$ approaches two different limiting values depending on the form of the turbulence. Ultraquantum or Vinen turbulence is the simplest form of quantum turbulence, where (in three-dimensions) there is a nearly random tangle with an apparent lack of large-scale motions in the velocity field. In contrast in the quasi-classical regime there is some structure to the quantised vortices, and at large length-scales (i.e. much larger than the typical intervortex spacing) non-zere course grained velocity and vorticity fields exist, and one would expect that these course grained fields are continuous functions and so a classical-like description becomes possible.  For these two different forms of quantum turbulence in the limit of zero temperature, it has been found that $\nu'_{\rm UQ}/\kappa  \sim \mathcal{O}(0.1)$ for the ultraquantum regime and $\nu'_{\rm QC}/\kappa  \approx \mathcal{O}(0.01-0.1)$ for the quasi-classical regime  \cite{Walmsley}.  

Within the context of the GPE, it has been estimated that $ \nu'/\kappa \sim \mathcal{O}(0.1)$ for three-dimensional ultraquantum turbulence \cite{Stagg} and $\nu'/\kappa \sim \mathcal{O}(1)$ for three-dimensional quasi-classical turbulence in a superfluid boundary layer \cite{Stagg2017}.  The larger values of $\nu'$ within GPE over superfluid Helium has been attributed to the fact that the vortices are many orders of magnitude closer to each other (relative to their core size) in GPE simulations.   Our value $\nu'/\kappa \approx 0.015$ is clearly much lower than these previous GPE-based results, despite comparable intervortex distances.  This leads us to conclude that the difference is due to the dimensionality of the flow.   To our knowledge, this is the first estimate of the effective viscosity for a 2D superfluid, and we hope that future studies will provide comparatives estimates of this quantity. 

\begin{figure}
\centering
\includegraphics[width=0.49\linewidth]{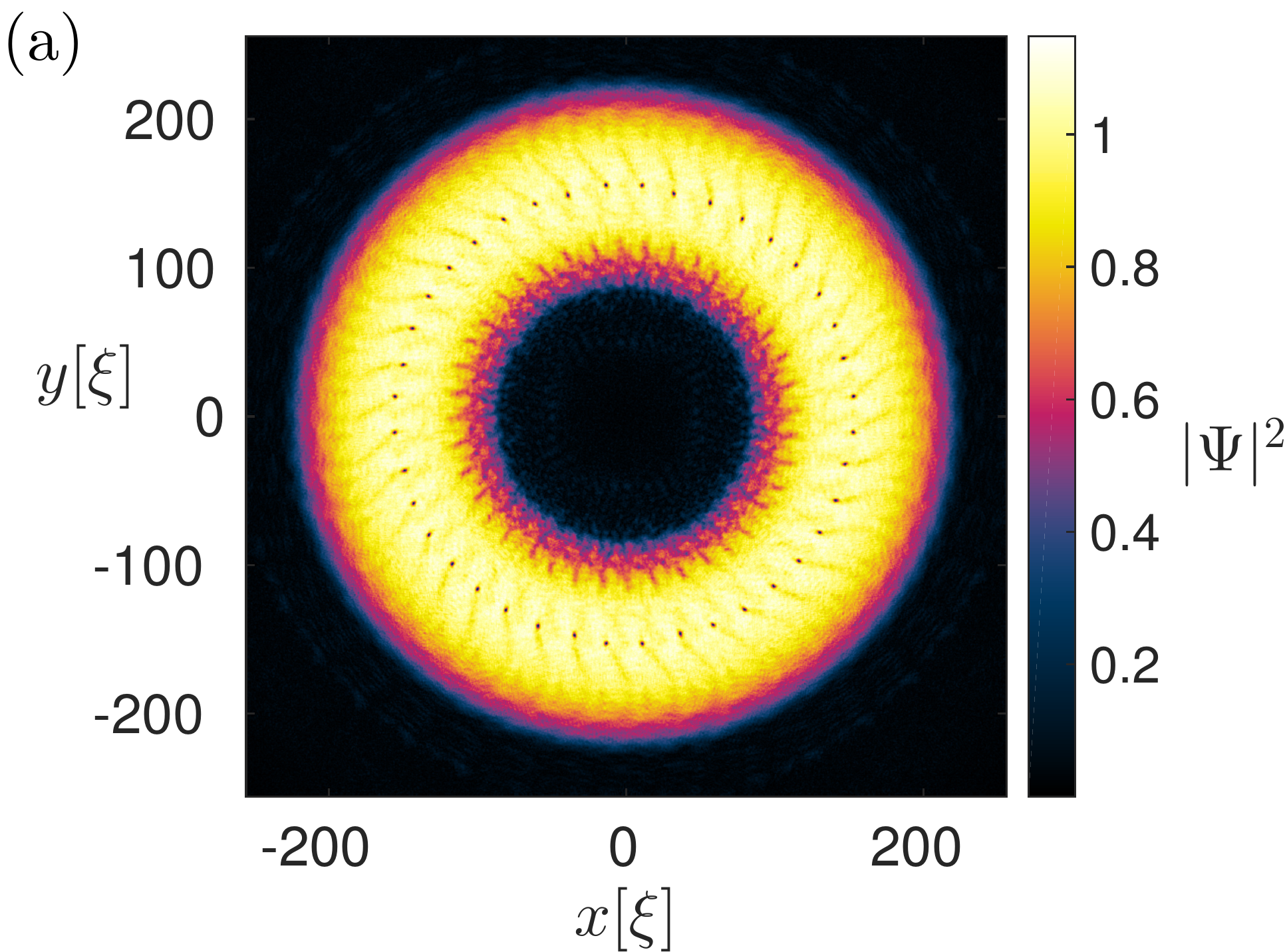}\hfill
\includegraphics[width=0.49\linewidth]{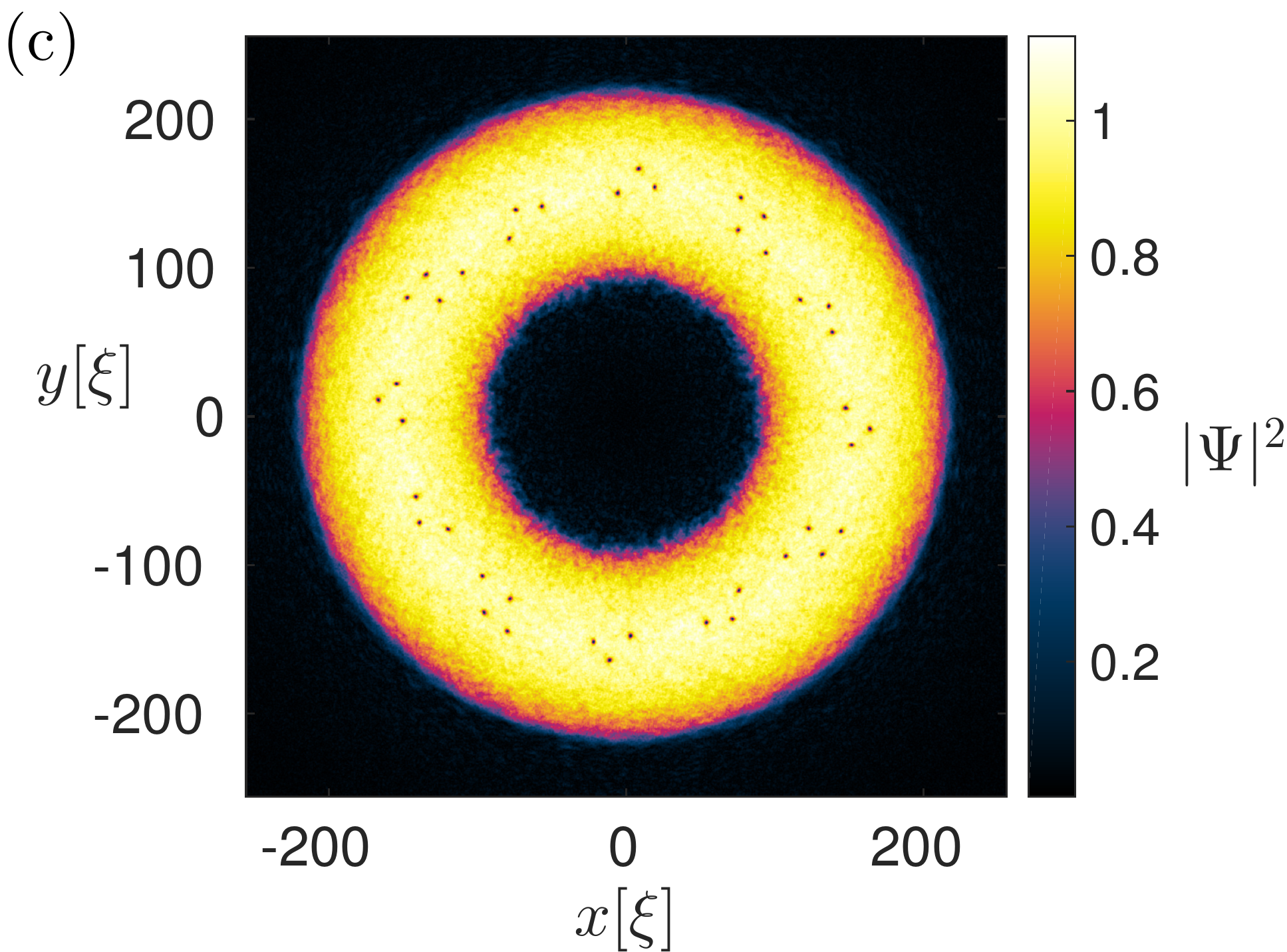}\\
\includegraphics[width=0.49\linewidth]{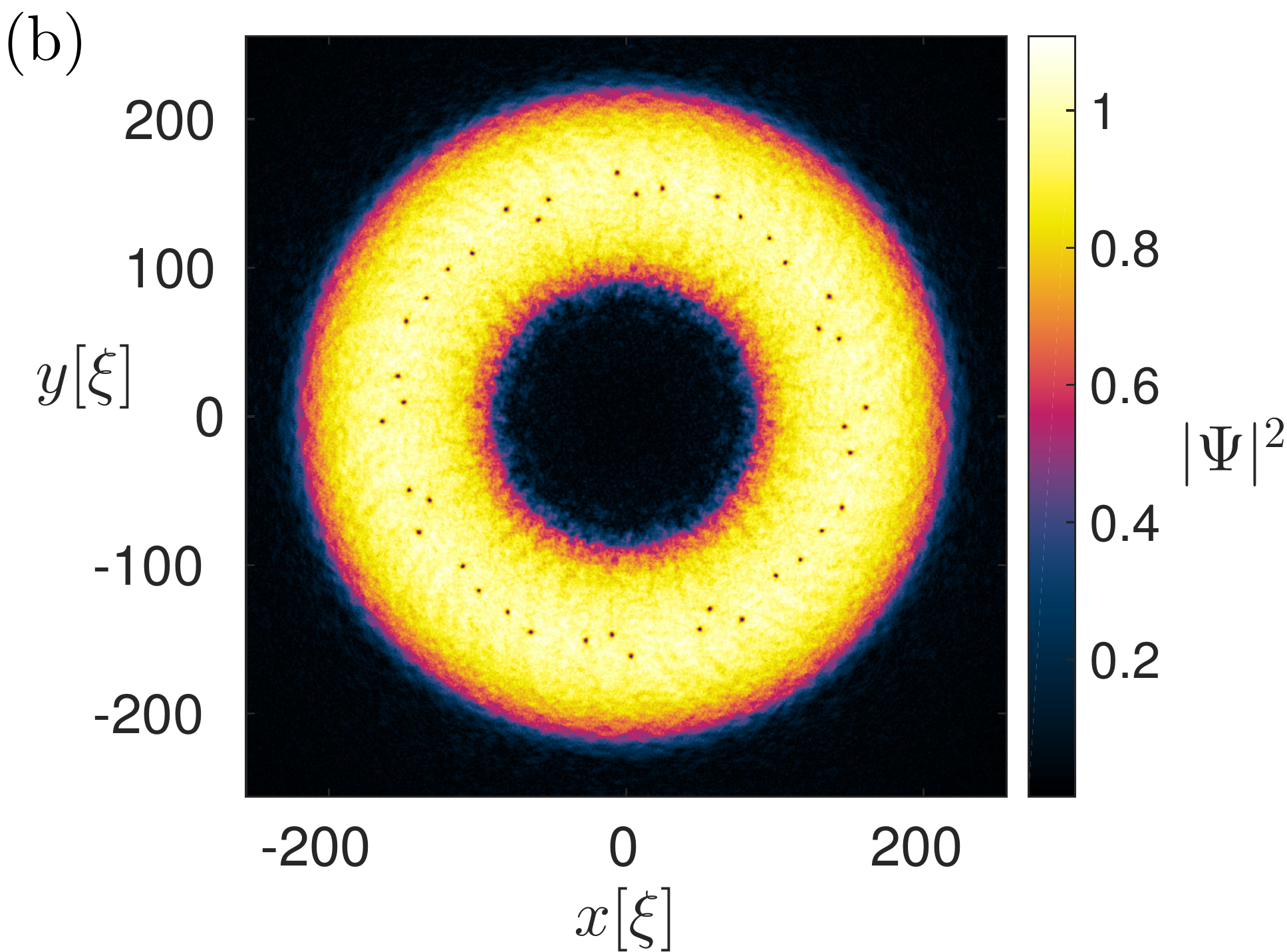}\hfill
\includegraphics[width=0.49\linewidth]{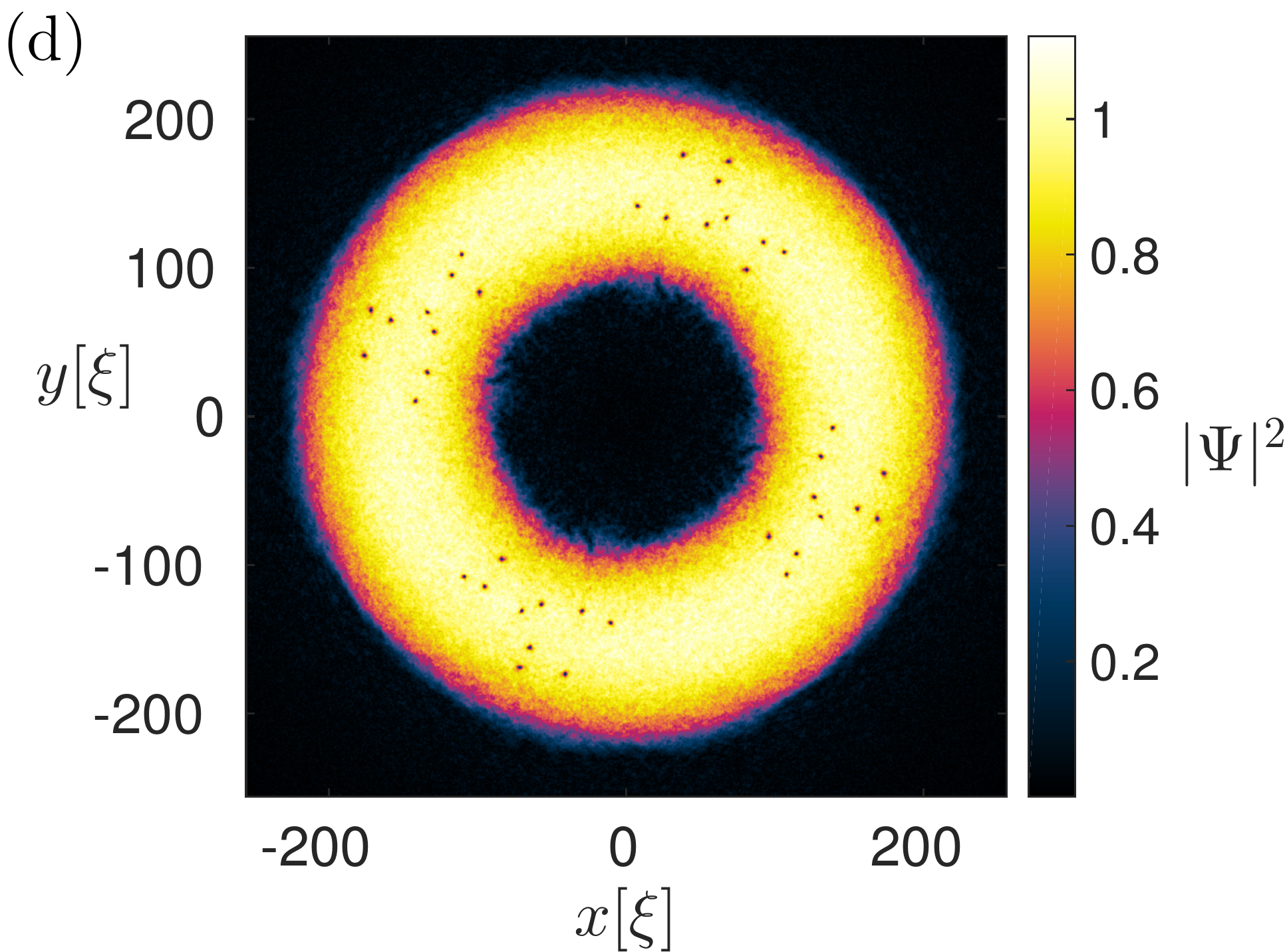}
\caption{Snapshots of the evolution of the density with $\mathcal{W}=20$, in a circular channel, with a small amount of noise added to the initial condition, for (a) $t=400$; (b) $t=800$; (c) $t=1200$; (d) $t=2500$. Note the KH instability is observed with an experimentally feasible system.  }
\label{fig8}
\end{figure}

Before we close we turn to an experimentally-feasible means to realize the KH instability in an atomic BEC.  A ring-trap geometry provides a natural setup to replicate our periodic channel, motivated by the experimental use of ring traps to study the superfluid dynamics of atomic BECs \cite{rings}.   In one such experiment, a Laguerre-Gauss beam was used to controllably impart angular momentum to the atoms, which served to phase imprint winding numbers up to $10$ \cite{Moulder2012}.   Figure \ref{fig8} shows the dynamics when the condensate is now confined to a ring-shaped channel  (simulated in a square domain, $D_x=D_y=256$).  As in the straight channel simulations, we impose counter-propagating flows in the outer/inner halves of the channel, and use a narrow barrier to initially separate the flows.  Following removal of the barrier, we see the establishment of a line of vortices which proceed to `roll-up' in a qualitatively similar manner to the simulation presented in Fig.~\ref{fig4}.   Note that it may be more convenient in practice to create the relative flow by initially phase imprinting the outer half of the ring-shaped channel while keeping the inner half in shadow (and thus stationary) by means of an optical mask.  Note also that our method of estimating the effective viscosity is experimentally achievable.  While it is not possible to directly measure the fluid velocity, it is now possible to experimentally identify both the positions and the circulations of the vortices \cite{Powiss2014,Seo2017,Johnstone2018}.  From this information the velocity field, and hence the coarse-grained momentum across the channel, can be readily reconstructed.

\section*{Conclusions}
In conclusion, we have demonstrated the analog of the famous classical Kelvin-Helmholtz instability in an atomic superfluid gas.  Two adjacent regions of the fluids which are initially in relative motion entrap a line of quantized vortices along their interface.  This quantum vortex sheet is unstable, and rolls up into small clusters of same-sign vortices.  Over time these clusters merge to create larger clusters.  When coarse-grained this flow mimicks a classical shear flow, allowing an effective viscosity to be estimated.  Once the cluster size becomes comparable to the channel width, secondary vortices of opposite sign become nucleated, mixing into the turbulent flow, and the end state is the segregation of the vortices into clusters of like-sign vortices.  These dynamics are experimentally accessible within ring-trapped atomic BECs.

\section*{Acknowledgements}

N.P. acknowledges support by the Engineering and Physical Sciences Research Council (Grant No. EP/R005192/1).

\end{document}